\documentclass{aa}  

\usepackage{graphicx}
\usepackage{txfonts}
\usepackage{graphicx}
\usepackage{natbib}
\usepackage{textcomp}

\def\nodata{ ~$\cdots$~ }

\begin{document} 

\title{High frequency breaks in optical active galactic nuclei power spectral density}

\author{Heechan Yuk\inst{1}
        \and
        Xinyu Dai\inst{1}
        }

\institute{University of Oklahoma, 
          440 W. Brooks St., Norman, OK 73019, USA
          }

\date{February 00, 2025}

\abstract
{Variability is a ubiquitous feature of active galactic nuclei (AGNs), and the characterisation of the variability is crucial to constrain its physical mechanism and proper applications in AGN studies. The advent of all-sky and high-cadence optical surveys allows more accurate measurements of AGNs variability down to short timescales and direct comparisons with X-ray variability from the same sample of sources.}
{We aim to analyse the optical power spectral density (PSDs) of AGNs with measured X-ray PSDs.}
{We use light curves from the All-Sky Automated Survey for SuperNovae (ASAS-SN) and the Transiting Exoplanet Survey Satellite (TESS) and use the Lomb-Scargle periodogram to obtain PSDs. The joint optical PSD is measured over up to six orders of magnitude in frequency space from timescales of minutes to a decade. We fit either a damped random walk (DRW) or a broken power law (BPL) model to constrain the PSD model and break frequency.}
{We find a set of break frequencies ($\lesssim10^{-2}$ day$^{-1}$) from DRW and BPL fits that generally confirm previously reported correlations between break frequencies and the black hole mass. In addition, we find a second set of break frequencies at higher frequencies ($>10^{-2}$ day$^{-1}$). We observe a potential weak correlation between the high-frequency breaks with the X-ray break frequencies and black hole mass.
We further explore the dependence of the correlations on other AGN parameters, finding that adding either X-ray, optical, or bolometric luminosity as the third correlation parameter can substantially improve the correlation significances. The newly identified high-frequency optical breaks can constrain different aspects of the physics of active galactic nuclei.
}
{}


\maketitle

\section{Introduction}
Variability is a ubiquitous characteristic of active galactic nuclei (AGNs), and is observed at all wavelengths (e.g., \citealt{mushotzky93}; \citealt{vandenberk04}) and on all timescales (e.g., \citealt{grandi92}; \citealt{markowitz03}). The mechanism driving the variability is still debated. A number of theories have been suggested, such as accretion disc instabilities (e.g., \citealt{kawaguchi98}; \citealt{trevese02}), global inhomogeneities from infall or thermal processes (e.g., \citealt{lightman74}; \citealt{shakura76}; \citealt{kelly09}), local disc communication processes on sound crossing or Alfven timescales \citep{dexter11}, magnetorotational instabilities (e.g., \citealt{balbus91}; \citealt{reynolds09}), and reprocessing between the disc and the corona on light crossing timescales (e.g., \citealt{haardt91}; \citealt{mchardy18}).

Though its origin is uncertain, AGN variability provides clues about the physical properties of the central engine. The variability at different wavelengths is known to arise from different regions of the AGN, and the variability time lags in different wavelengths can be used for reverberation mapping to measure the physical size of the AGN structure and the mass of the central supermassive black hole (SMBH) (e.g., \citealt{wandel99}; \citealt{kaspi00}; \citealt{peterson05}). Different classes of AGN also show different variability features (e.g., \citealt{ptak98}), and by studying these differences, we can learn specific details about each class. As a behaviour common to all AGN, variability can also be used to find AGN (e.g., \citealt{butler11}; \citealt{macleod11}; \citealt{ruan12}; \citealt{treiber22}; \citealt{yuk22}).

The stochastic variability of AGN can be characterised by its power spectral density (PSD). The PSDs are characterised by red noise, where the power increases for lower frequencies and then saturates below some frequency (e.g., \citealt{uttley02}). For optical variability this is commonly modelled as a damped random walk (DRW) (e.g., \citealt{kelly09}; \citealt{kozlowski16}). In these models, the DRW break frequency, the frequency where the PSD starts to flatten, is correlated with the mass of the SMBH powering the AGN (e.g., \citealt{kelly09}; \citealt{macleod10}; \citealt{burke21}).
With the recent advent of high-cadence optical data, we can now probe the optical PSD on short timescales (e.g., \citealt{smith18}), where steeper power law slopes than DRW were inferred \citep{mushotzky11}.

The X-ray variability of AGN is observed down to very short timescales ($<$ day) (e.g., \citealt{grandi92}; \citealt{uttley02}), with many PSD and break frequency studies (e.g., \citealt{uttley02}; \citealt{mchardy04}; \citealt{gonzalezmartin12}; \citealt{gonzalezmartin18}; \citealt{smith18}). 
However, previous studies are limited by the differences in optical and X-ray timing surveys, and the direct comparison between optical and X-ray PSDs for the same set of sources were difficult.
For this study, we combine the long-term, low-cadence, ground-based data from the All-Sky Automated Survey for SuperNovae (ASAS-SN; \citealt{shappee14}; \citealt{kochanek17}) and the short-term, high-cadence, space-based data from the Transiting Exoplanet Survey Satellite (TESS; \citealt{ricker15}). We construct a PSD separately for each data set and then merge them to identify the break frequencies. 
This allows the first systematic comparison of optical and X-ray break frequencies in the power spectra of AGN.
For this comparison, we use the sample of AGN with X-ray break frequencies measured by \citet{gonzalezmartin18}.
We introduce the sample, the ASAS-SN and TESS data, and the data analysis procedure in Section \ref{secdata}. We describe the methods for extracting and fitting the PSDs in Section \ref{secmethods}. In Section \ref{secresults}, we present the resulting break frequencies and examine how they relate to previous results. We discuss the implications in Section \ref{secdiscussion}.

\section{Data}
\label{secdata}

\subsection{Data properties and the sample}
ASAS-SN (\citealt{shappee14}; \citealt{kochanek17}) began surveying the sky in 2012. After several expansions, it now consists of 20 14-cm telescopes distributed over five ``units" and can cover the visible sky nightly in good conditions. The two original units initially used the $V$-band filter. The three new units commissioned in 2017 use $g$-band filters and the original two units switched to $g$-band in 2018. During the transition, there were approximately 400 days when both $V$-band and $g$-band were used. ASAS-SN images have limiting magnitudes of $V\sim16.5-17.5$ and $g\sim17.5-18.5$ depending on lunation, a field of view of 4.5 deg$^2$, a pixel scale of 8\farcs0, and a typical FWHM of $\sim2$ pixels.

Launched in 2018, TESS (\citealt{ricker15}) is equipped with four cameras and its TESS bandpass spans a wavelength range of 600 to 1000 nm. Each camera has a field of view of 24\textdegree\, by 24\textdegree. The combined 24\textdegree\, by 96\textdegree\, field of view is called a sector. TESS observes each sector for approximately 27.4 days. For the primary mission, the 26 sectors observed until July 2020, the full-frame images (FFIs) were taken with a 30-minute cadence. The first extended mission, which lasted until sector 55 in September 2022, collected FFIs with a 10-minute cadence. In the current, second extended mission, the FFIs are taken with a 200-second cadence. Each sector is divided into two cycles, each lasting for approximately 13 days, with $\sim$1 day gap in between for data transmission. 
The FFIs have the pixel scale of 21\farcs0 with a typical FWHM of $\sim2$ pixels.
The FFIs used for this study can be found in Mikulski Archive for Space Telescopes (MAST\footnote{http://dx.doi.org/10.17909/0cp4-2j79}).

\citet{gonzalezmartin18} provided the most recent compilation of AGN with measured X-ray breaks in their power spectra.
This sample includes 22 AGNs, with their black hole masses, $\log (M_{\textrm{BH}}/M_{\odot})$, ranging from 5.4 to 8.5. Majority of the black hole masses are measured by reverberation mapping, while others are measured by different methods including $M-\sigma$ relation, 5100\AA\hspace{1pt} continuum luminosity, water masers, and radio fundamental plane. Their bolometric luminosities, $\log (L_{\textrm{bol}}/\textrm{erg s}^{-1})$, range from 41 to 46. These AGNs are nearby, with their redshifts ranging from 0.001 to 0.05, with the exception of PKS 0558-504 with a redshift of 0.14.
These AGNs all have \textit{XMM-Newton} observations with 88--1095 ks exposure times. \citet{gonzalezmartin18} binned the 2--10 keV data into 50-second bins (with a few exceptions using 100- and 200-second bins), and produced light curve segments of 10, 20, 40, 60, 80, and 100 ks. The PSD of each light curve was modelled with a broken power law to measure the break frequency. Each light curve segment for a target was fit separately, leading to a range of break frequency estimates. We compare the optical and X-ray properties of the power spectra of this sample in this paper.
Five of these targets (MRK 335, IC 4329A, NGC 5506, ARK 564, and NGC 7469) were not observed by TESS by the end of the first extended mission. For these objects, we use only the ASAS-SN light curves.

\subsection{Data reduction}
To extract the ASAS-SN light curves, we used image subtraction \citep{alard98, alard00} and aperture photometry (see \citealp{jayasinghe18} for more details). For calibration, we used the AAVSO Photometric All-Sky Survey (APASS) catalogue \citep{henden15}. We corrected the zero-point offsets between different cameras and recalculated the photometric errors as described in \citet{jayasinghe18} and \citet{jayasinghe19}, respectively.

We used a similar method to extract the TESS light curves, with the modifications described in \citet{fausnaugh21} and \citet{vallely21}. We first cut out 750-pixel postage stamps around the target from the FFIs. Then we create a reference image using 100 high-quality images that have no mission-provided data quality flags and have point spread function (PSF) widths and background levels lower than the median values for the sector. We then scale and subtract the reference image from all the individual images. We use a median filtering method to minimise the effects of background structures (the ``straps" and complex internal reflection artefacts) and then perform PSF photometry on the subtracted images.
We use the subtraction method over aperture photometry because TESS's large pixel size makes source blending a significant challenge to aperture photometry, while the subtraction method simply measures the flux changes relative to the reference image \citep{vallely21}.

\begin{figure*}
    \centering
    \includegraphics[width=0.8\hsize]{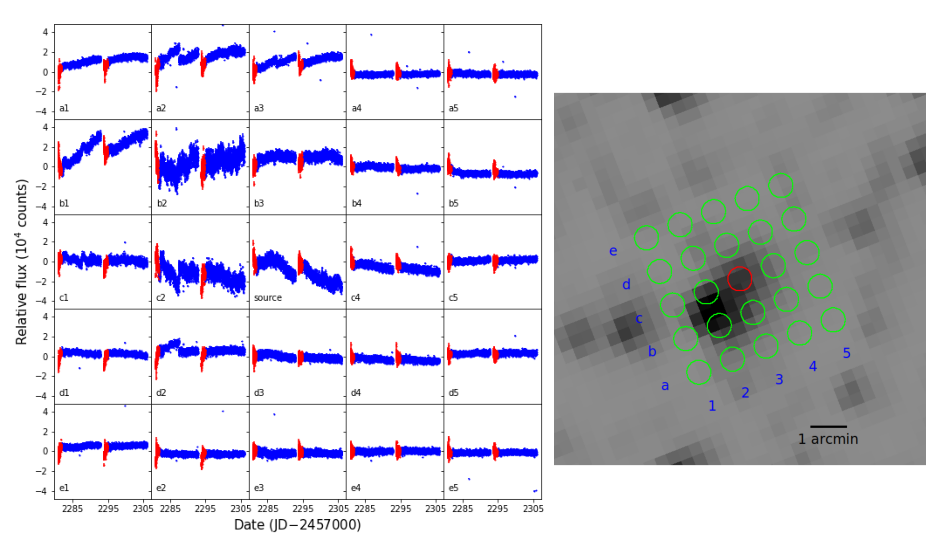}
    \caption{Left: The TESS sector 36 light curve of NGC 3783 and nearby pixels. The centre panel labelled ``source" is the light curve of the target and the other 24 panels are the light curves of nearby pixels. The red points are the data points with uncertainties greater than the standard deviation of uncertainties from the median of each light curve. Right: The TESS image and the locations of NGC 3783 (red circle) and the nearby pixels (green circles) where the light curves on the left were extracted.}
    \label{ngc3783tesslc}
\end{figure*}

Once the initial TESS light curve is extracted, a few additional adjustments are made. To check for any anomalies, we also extracted light curves for a grid of nearby pixels as shown in Figure \ref{ngc3783tesslc}. At the beginning of each cycle of a TESS Sector, there is often a spike in the backgrounds lasting for a few days. For some objects, this greatly affects the photometry. For example, Figure \ref{ngc3783tesslc} shows that all the nearby pixels display the same pattern at the beginning of each cycle. 
The affected segments tend to have greater flux uncertainty, so we computed the median and standard deviation of the flux uncertainty, and took out the portion with uncertainties that deviate significantly from the median.
The observations for a sector are interrupted at the middle for data transmission. Sometimes, there is an offset between these two sub-sections  (also seen in Figure~\ref{ngc3783tesslc}), which we also correct. 

\begin{figure*}
    \centering
    \includegraphics[width=0.8\hsize]{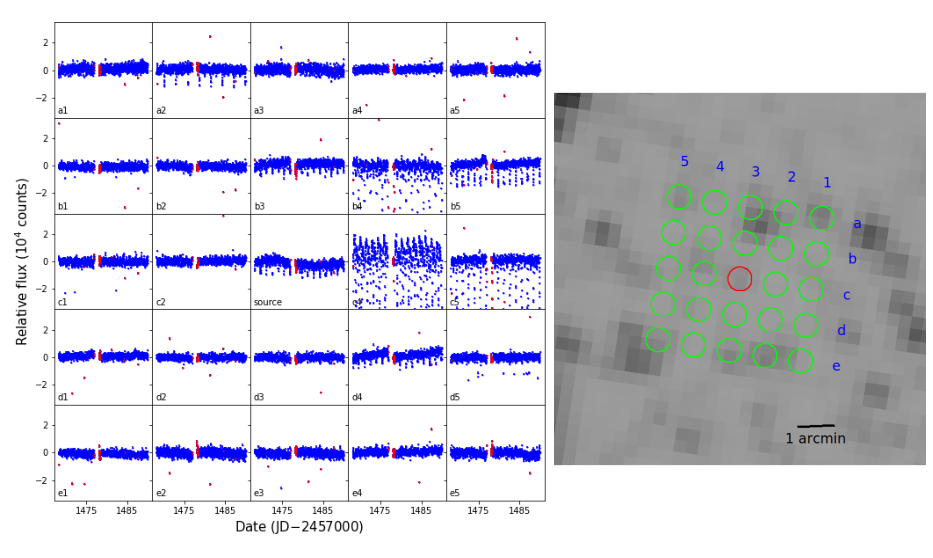}
    \caption{Same as Figure \ref{ngc3783tesslc} for the TESS sector 6 light curve of 1H 0707-495.  The source light curve shows periodic variability due to a nearby eclipsing binary, so this TESS light curve was discarded.}
    \label{1h0707-495tesslc}
\end{figure*}

Some sources have anomalous TESS light curves and are discarded. For example, the TESS light curves of 1H 0707$-$495 (Figure \ref{1h0707-495tesslc}) display a periodic behaviour where the flux drops regularly. However, this pattern is observed in most of the nearby pixels, and it is due to contamination from a nearby eclipsing binary. All six TESS sectors are affected by the binary, so we rejected all the TESS light curves of this target for analysis. We confirmed that there are no other similar cases by checking the ASAS-SN variable stars database (\citealt{shappee14}; \citealt{jayasinghe18}) for variable stars near each target. Similarly, when we observe unusual features in a TESS light curve, we look at the light curves of nearby pixels to determine if the feature is likely to be real or that the sector should be rejected. Table \ref{tesssecs} lists the TESS sectors used for each target and labels which ones are discarded.

\begin{table}
\caption{TESS sectors used for the sample.}
\label{tesssecs}
\centering
\begin{tabular}{l l}
\hline
Name         & TESS sectors \\
\hline \hline
MRK 335      & \nodata                                     \\
ESO 113-G010 & 1, 2, 28, 29                                \\
Faiall 9     & 2, 28, 29                                   \\
PKS 0558-504 & 1$^*$, 5, 6, 7, 11$^*$, 32, 33, 34          \\
1H 0707-495  & 6$^*$, 7$^*$, 8$^*$, 32$^*$, 33$^*$, 34$^*$ \\
ESO 434-G040 & 9, 35                                       \\
NGC 3227     & 45, 46, 48                                  \\
RE J1034+396 & 21, 48                                      \\
NGC 3516     & 14, 20, 21, 41, 47, 48                      \\
NGC 3783     & 10, 36, 37                                  \\
NGC 4051     & 22, 49                                      \\
NGC 4151     & 49                                          \\
MRK 766      & 22                                          \\
NGC 4395     & 22, 49                                      \\
MCG-06-30-15 & 11$^*$, 37                                  \\
IC 4329A     & \nodata                                     \\
Circinus     & 11$^*$, 12$^*$, 38                          \\
NGC 5506     & \nodata                                     \\
NGC 5548     & 23, 50                                      \\
NGC 6860     & 13, 27                                      \\
ARK 564      & \nodata                                     \\
NGC 7469     & \nodata                                     \\
\hline
\end{tabular}
\end{table}

The ASAS-SN light curves are measured in mJy or magnitudes, while the TESS light curves are measured in units of counts relative to the reference frame. Before we extract the power spectra, we converted the TESS light curves into mJy, using the TESS calibration that one count $=20.44\pm0.05$ mag with a zero point of 2631.9 Jy \citep{vanderspek18}.

\begin{figure*}
    \centering
    \includegraphics[width=0.8\hsize]{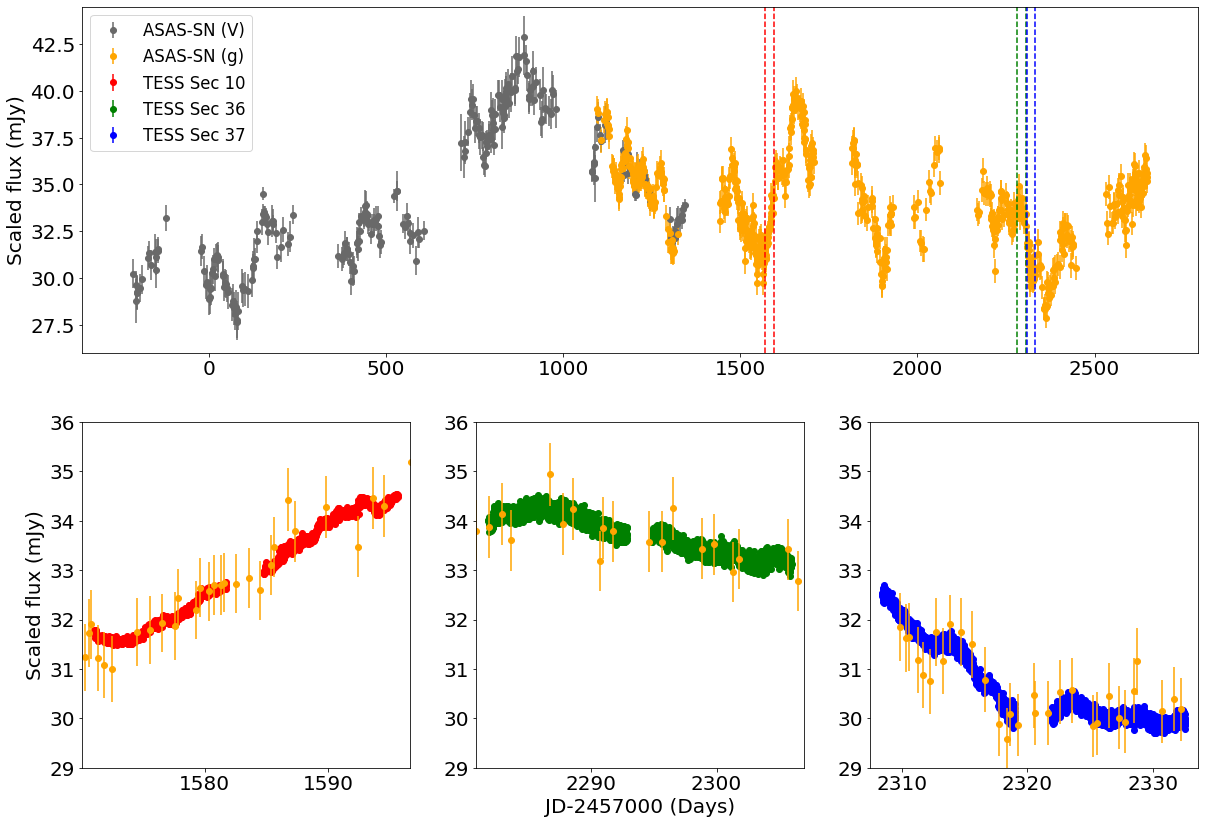}
    \caption{Top: The ASAS-SN light curves of NGC 3783 with the TESS sectors marked. Bottom: TESS light curves along with ASAS-SN data. The light curves are scaled to overlap.}
    \label{ngc3783lc}
\end{figure*}

\begin{figure*}
    \centering
    \includegraphics[width=0.8\hsize]{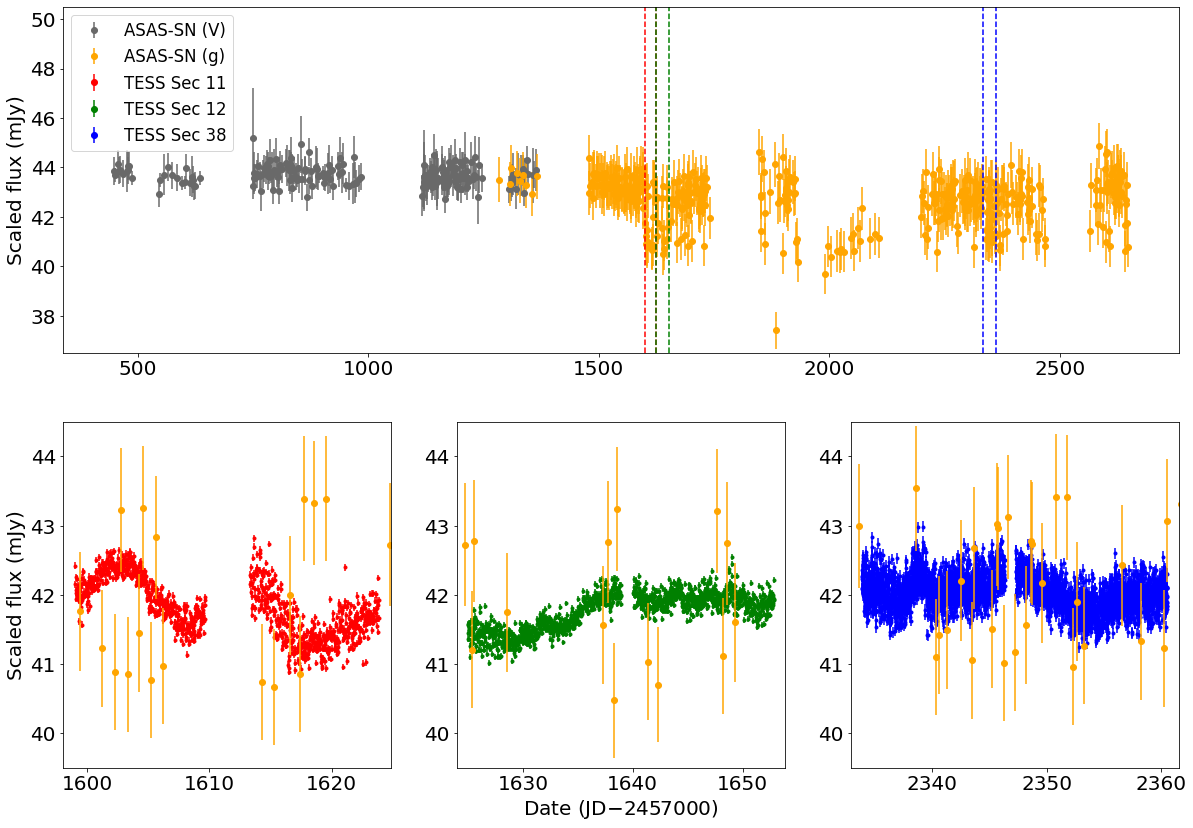}
    \caption{Same as Figure \ref{ngc3783lc} for Circinus with TESS sectors 11, 12, and 38.}
    \label{circinuslc}
\end{figure*}

A representative compilation of light curves (NGC 3783) is shown in Figure \ref{ngc3783lc}. It shows that the TESS and ASAS-SN light curves have similar trends with offsets due to the filter differences. The light curves for Circinus are shown in Figure \ref{circinuslc}. For sectors 11 and 12, a sinusoidal pattern was found in the nearby pixels affecting the target's light curve, while that pattern is not seen in sector 38. We discarded sectors 11 and 12 and used only sector 38 for the PSD analysis.

\section{Methods}
\label{secmethods}

\subsection{Power spectra}
In general, there are two approaches to constrain PSDs from the light curves: one is the backward method of transforming the light curves into frequency domain and the other is the forward method of assuming a parametric model to fit the light curves. Most previous studies of optical PSDs of AGNs were using the forward method because of the sparse sampling of the light curves (e.g., \citealt{kelly09}; \citealt{kozlowski10}; \citealt{macleod10}; \citealt{zu13}); however, with the high cadence optical survey, the more intuitive backward method were used in more analyses (e.g., \citealt{mushotzky11}; \citealt{burke20}; \citealt{petrecca24}). 
Since the forward method can only apply to parametrized PSD models, mostly the damped random walk model in optical studies, it is beneficial to use the backward methods to infer the general characteristics of the PSD with the inclusion of the new high-frequency TESS data in this paper, which potentially can reveal new PSD features that are absent in commonly assumed optical PSD models.  In addition, most previous forward modelling analyses were applied to one dataset, and the computation time and calibration effort will increase significantly when dealing with multiple datasets as in this study.
Here, we used a backward Lomb-Scargle periodogram (\citealt{lomb76}; \citealt{scargle82}) to construct the power spectra with simulations to constrain its biases imposed in the PSD break measurement. A forward-fitting analysis will be presented in a separate paper.

The lower frequency limit is set at $1/T$, where $T=t_N-t_1$ is the time span of the observations, and the upper limit is set at $1/(2\Delta T_{\textrm{samp}})$, where $\Delta T_{\textrm{samp}}$ is the typical sampling interval (1 day for ASAS-SN, 30 minutes for TESS sectors $\le$26, 10 minutes for TESS sectors $>$26).
The raw periodograms from the different data sets do not line up perfectly, so we apply several corrections. First, we apply the rms normalisation $2\Delta T_{\textrm{samp}}/(\bar{x}^2N)$ (e.g., \citealt{vaughan03}), where $\bar{x}$ is the mean flux. Such rms normalisations will also minimise the biases due to the filter differences of the different datasets, assuming that the wavelength-dependent AGN variability mainly manifests in the rms amplitudes (e.g., \citealt{macleod10}). We next separately normalised the PSDs between the ASAS-SN $V$- and $g$-band PSDs, and between the multiple TESS sectors' PSDs. There can be additional offsets between ASAS-SN and TESS PSD measurements such as differences in the rms variability between wavelengths. It is challenging to normalise the ASAS-SN and TESS PSDs because the ASAS-SN PSDs are dominated by noise in the overlapping region.
So, we preliminarily normalise the TESS and ASAS-SN components by lining up the PSDs at $<10^{-2}$ day$^{-1}$. Then, for the fitting process, we include the normalisation factor for each PSD as free parameters.

\begin{figure*}
    \centering
    \includegraphics[width=0.8\hsize]{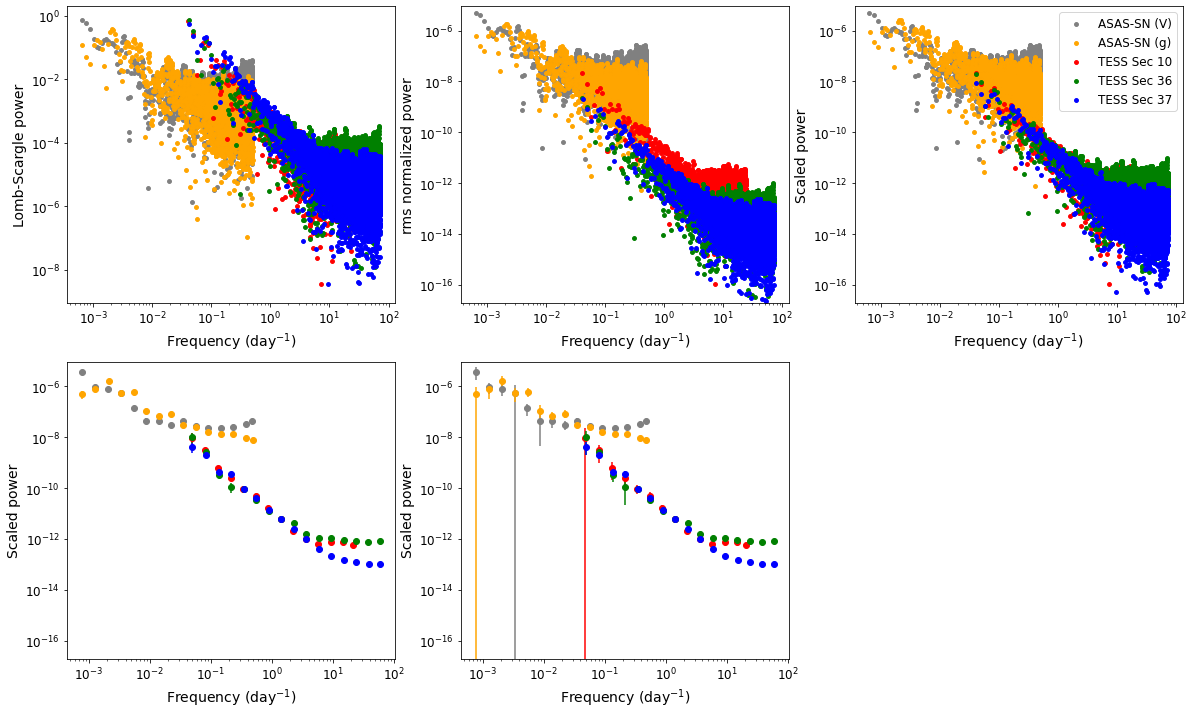}
    \caption{Steps to produce the PSDs used in the model fits. Top: The original Lomb-Scargle periodogram of NGC 3783 (left), with rms normalisation (middle), and after additional scaling to line up the PSDs (right). Bottom: The binned periodogram with the standard deviation of the mean as the uncertainty (left), and with the additional estimated systematic uncertainties (middle).}
    \label{ngc3783psd}
\end{figure*}

Before fitting, the periodogram is binned and averaged linearly into logarithmic frequency bins with a width of a factor of 1.6 (0.2 dex) in frequency. The uncertainties are the standard deviations about the mean. To estimate any additional systematic error contribution, we fit a simple power law plus white noise model, $P(\nu)=A\nu^{-\alpha}+C$, calculated the reduced $\chi^2$ statistic, $\chi^2_{\rm{red}}$, and then scaled the uncertainties to make $\chi^2_{\rm{red}}$ unity. On average, $\chi^2_{\rm{red}}$ before accounting for systematic uncertainties is about 5 for the TESS data, 25 for the V-band ASAS-SN data, and 10 for the g-band ASAS-SN data. Figure \ref{ngc3783psd} shows an example of the periodograms before and after the scaling, binning, and the addition of the estimated systematic uncertainties.

\subsection{Fitting}
\begin{figure}
    \centering
    \includegraphics[width=\hsize]{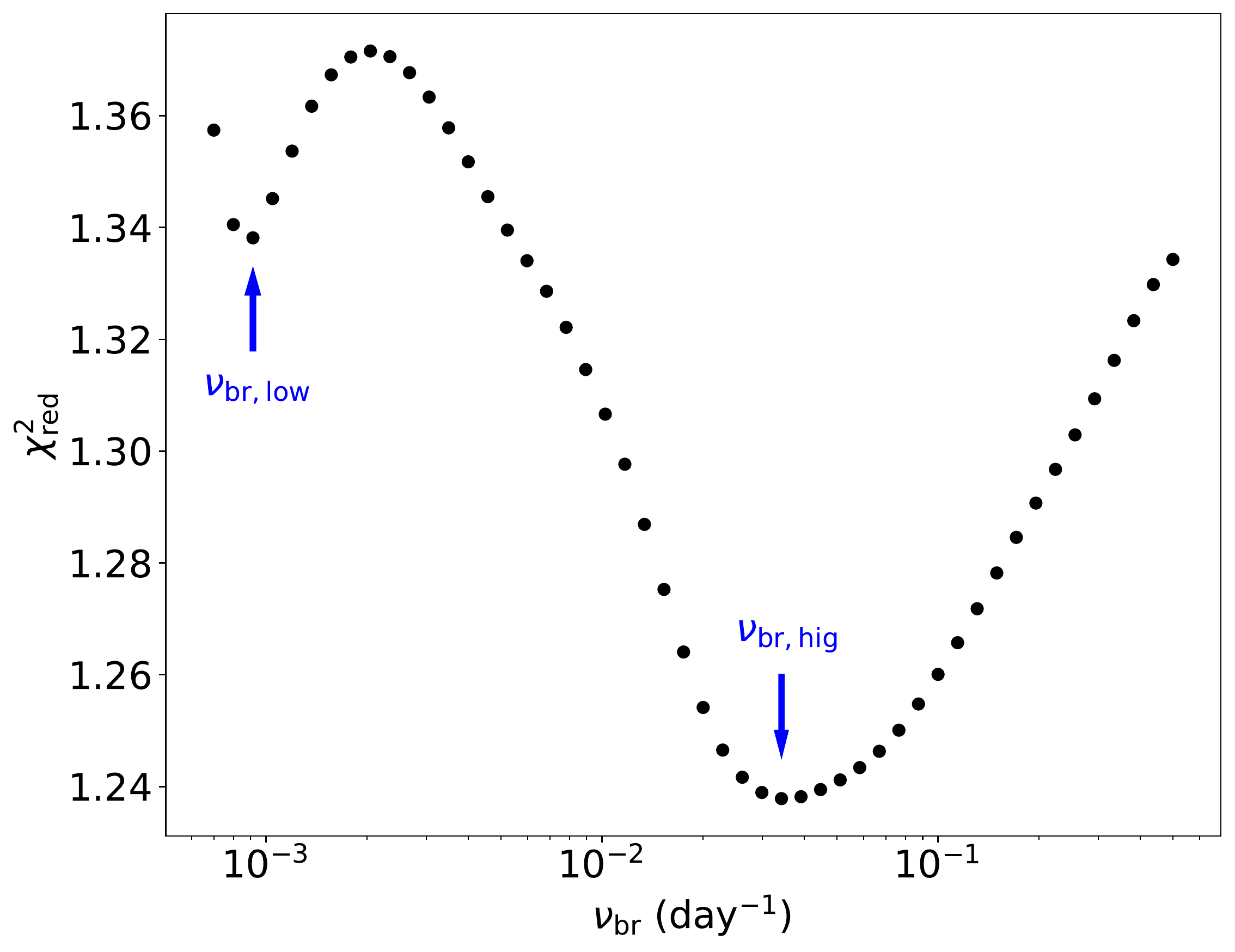}
    \caption{The reduced chi-squared for the broken power law model fitting as a function of the break frequencies for MRK 766. The positions we identify as the two break frequencies are labelled.}
    \label{redchisq_2valleys}
\end{figure}

We fit two models to the periodogram, the broken power law and damped random walk (DRW) models. The main difference between the two models is that the low-frequency and high-frequency power indices are not fixed at 0 and $-2$, respectively, in the broken power law model. We fit all models by minimising a $\chi^2$ statistic. We first fit a broken power law plus white noise model. Compared to the DRW model, this model has several additional parameters, so we approach the fit in several steps. First, we fit the simple power-law with white noise model, $P(\nu)=A\nu^{-\alpha}+C$ to the individual data sets. Using those fit parameters as initial guesses, the combined periodograms are fit with the smoothly broken power-law plus white noise power spectrum,

\begin{equation}
    P(\nu) = A_i \bigg(\frac{\nu}{\nu_{\textrm{br}}}\bigg)^{-\alpha_1}
    \bigg\{\frac{1}{2}\bigg[1+\bigg(\frac{\nu}{\nu_{\textrm{br}}}\bigg)^{1/\Delta}\bigg]\bigg\}^{(\alpha_1-\alpha_2)\Delta}+C_i,
\end{equation}
where $\nu_{\textrm{br}}$ is the break frequency, $A_i$ is the amplitude at $\nu_{\textrm{br}}$, $\alpha_1$ and $\alpha_2$ are the low and high frequency slopes, respectively, $\Delta$ is the smoothness parameter, where the larger values of $\Delta$ mean smoother break, and $C_i$ is the noise level for light curve $i$ \citep{astropy22}. The parameters $\nu_{\textrm{br}}$, $\alpha_1$ $\alpha_2$, and $\Delta$ are the same for all PSDs, while the amplitude $A_i$ and noise $C_i$ are different for each data set. The adjustments in the amplitudes allow us to account for any remaining normalisation differences between different PSD measurements. We first made fits with $\Delta$ fixed at 0.1, 0.2, 0.3, 0.4, and 0.5 and all other parameters free. We found that $\Delta=0.1$ resulted in the smallest $\chi^2$, so we fixed $\Delta=0.1$ for all subsequent fits.
Next, we test the dependence of $\chi^2$ on two model parameters: the TESS-to-ASAS-SN scaling factor and $\nu_{\textrm{br}}$. Though we adjust the $A_i$ individually, tests showed that the results are very sensitive to the initial conditions. So, we fit the data on a fixed 2-dimensional grid of scaling factor and $\nu_{\textrm{br}}$. The scaling factor that yields the overall lowest $\chi^2$ was chosen and the $\nu_{\textrm{br}}$ at the $\chi^2$ minimum is used as the initial guess for the model, with a $\pm10\%$ prior constraint on the frequency. Because there are still many parameters to fit, we used the shared value of the amplitude. There are two $\chi^2$ minima in some cases, as shown by the example in Figure \ref{redchisq_2valleys}. For these, we separately tried fits at both minima. After this fit, the constant terms $C_i$ are constrained to be within $\pm10\%$ of the values from the previous fit and the amplitudes $A_i$ are allowed to vary independently. For the final fit, we constrained the amplitudes $A_i$ and noise terms $C_i$ to be within 10\% of the prior trial fit values and allowed the power law indices to be free parameters.

\begin{figure*}
    \centering
    \includegraphics[width=0.8\hsize]{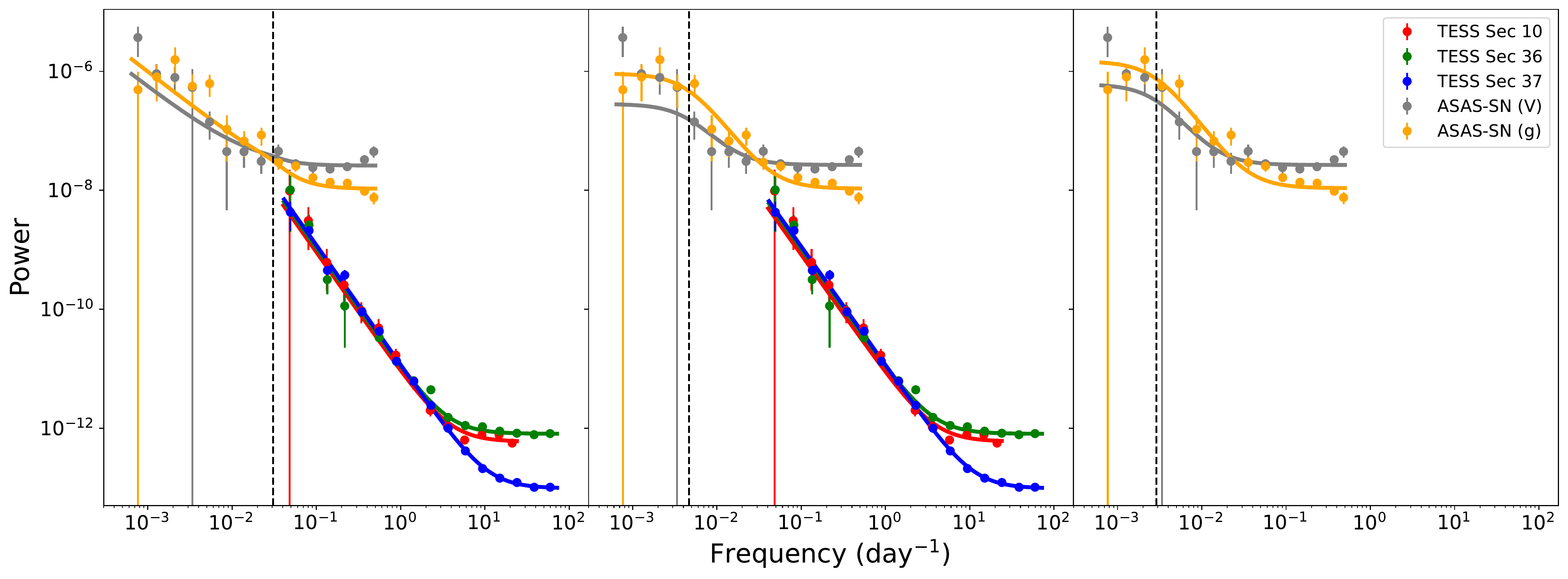}
    \caption{The binned periodogram of NGC 3783 fit with a broken power law with the high break frequency (left), a DRW using both TESS and ASAS-SN data (middle), and using only the ASAS-SN (right), The vertical dashed lines mark the break frequencies.}
    \label{ngc3783fit}
\end{figure*}

\begin{figure*}
    \centering
    \includegraphics[width=0.8\hsize]{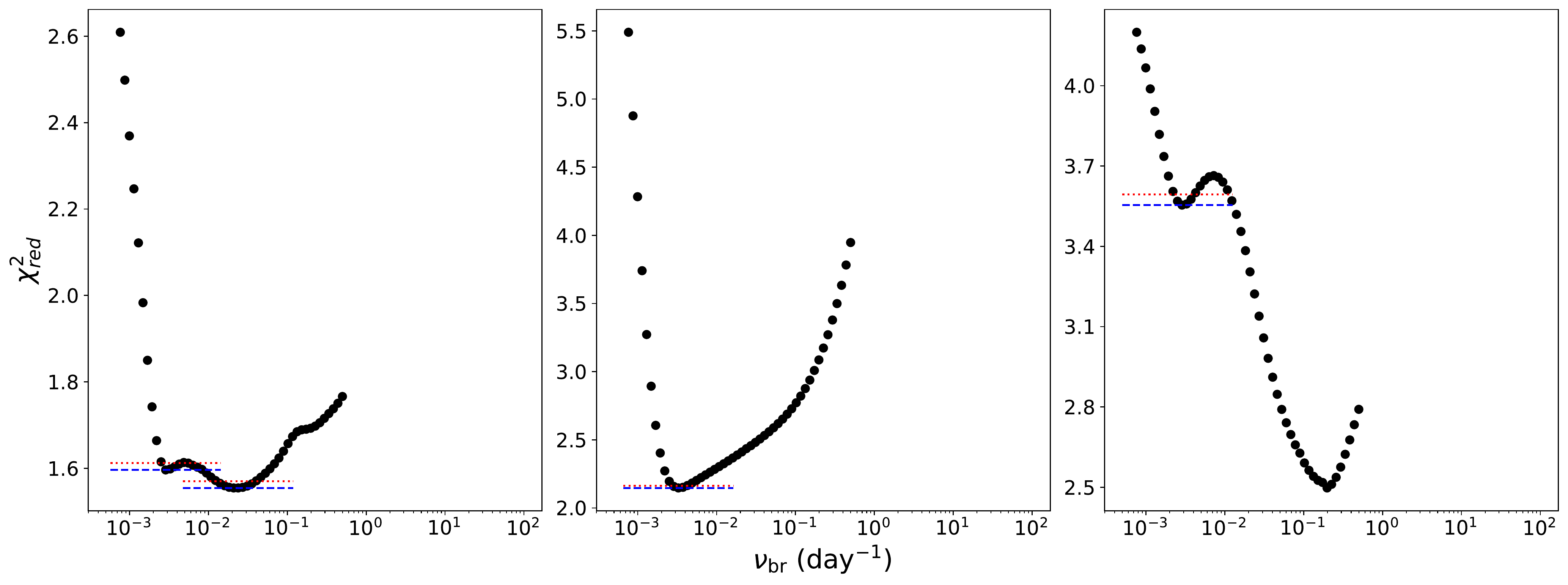}
    \caption{Reduced chi-squared as a function of break frequency for the three fits shown in Figure \ref{ngc3783fit}. There are two minima for the DRW fit to the ASAS-SN PSDs, but the minimum at the higher frequency is in the noise-dominated region, so the low-frequency break is used. The blue dashed lines indicate the local minima and the red dotted lines indicate the 1-sigma confidence level.}
    \label{ngc3783redchisq}
\end{figure*}

We also fit the DRW model, also known as the continuous-time first-order autoregressive (CAR(1)) process (e.g., \citealt{kelly09}), plus white noise. The DRW model has the power spectrum

\begin{equation}
    P(\nu) = \frac{2\sigma^2\tau^2}{1+(2\pi\tau\nu)^2},
\end{equation}
where $\sigma$ is the characteristic timescale and $\tau$ is the relaxation time. The relaxation time $\tau$ defines the break frequency of $\nu_{\textrm{br}}=1/2\pi\tau$. Since most previous optical PSD studies did not use light curves with a high cadence of TESS, we fit the DRW model to both the separate and combined ASAS-SN and TESS PSDs. In general, the DRW fits yield break frequencies in the ASAS-SN frequency regime, which is consistent with previous studies.
An example of the combined PSDs with the fits and the corresponding $\chi^2$ are shown in Figure \ref{ngc3783fit} and \ref{ngc3783redchisq}.

\subsection{Comparison between long-term ASAS-SN and TESS light curves and PSDs}
\begin{figure*}
    \centering
    \includegraphics[width=0.8\hsize]{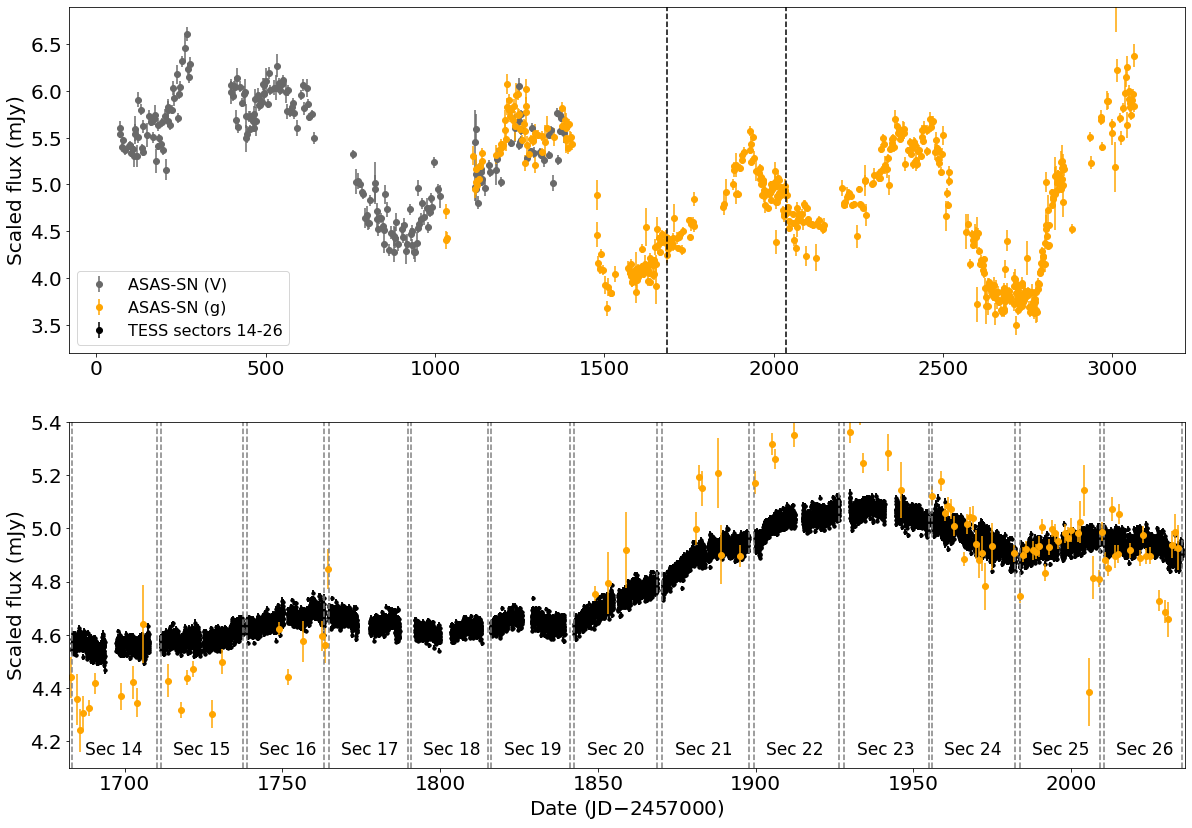}
    \caption{Top: The ASAS-SN light curve of PG 1613+658 with the observation window of PG 1613+658 during the TESS primary mission marked. Bottom: The TESS primary mission light curve of PG 1613+658 along with the ASAS-SN g-band data. The vertical dashed lines are the sector boundaries.}
    \label{pg1613+658lc}
\end{figure*}

Since most of the high-frequency breaks lie in the overlapping frequency range of the TESS and ASAS-SN PSDs, it is important to compare TESS and ASAS-SN light curves and PSDs on these time scales for potential calibration issues or variability characteristic differences between filters.  Typical TESS Sectors are too short for this purpose. However, TESS observes the ecliptic poles continuously for 13 sectors, so it is possible to extract long-term, nearly continuous TESS light curves for objects near the poles. For this analysis, we used PG~1613+658, a Seyfert I galaxy near the North ecliptic pole observed over Sectors 14 to 26 during the primary mission. The combined light curve is shown in Figure \ref{pg1613+658lc}.

\begin{figure}
    \centering
    \includegraphics[width=\hsize]{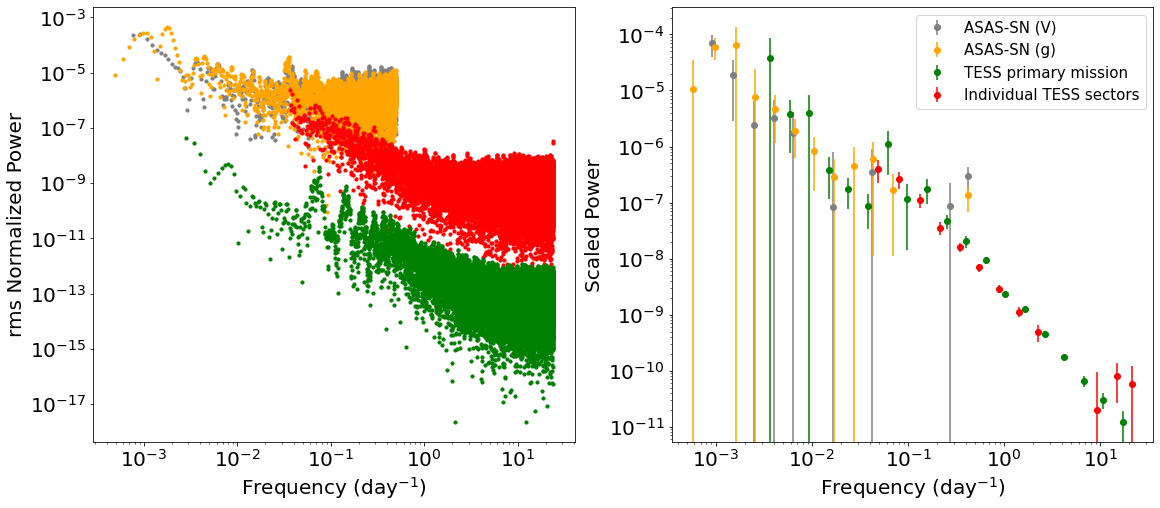}
    \caption{Left: The rms normalised PSDs of PG 1613+658. The PSD in green is constructed using the continuous 13 sector TESS light curve, while the red is the collection of the individual sector PSDs. Right: The binned PSDs with the white noise level subtracted and scaled to line up. The individual TESS sector PSDs are averaged before being binned.}
    \label{pg1613+658psd}
\end{figure}

\begin{figure}
    \centering
    \includegraphics[width=\hsize]{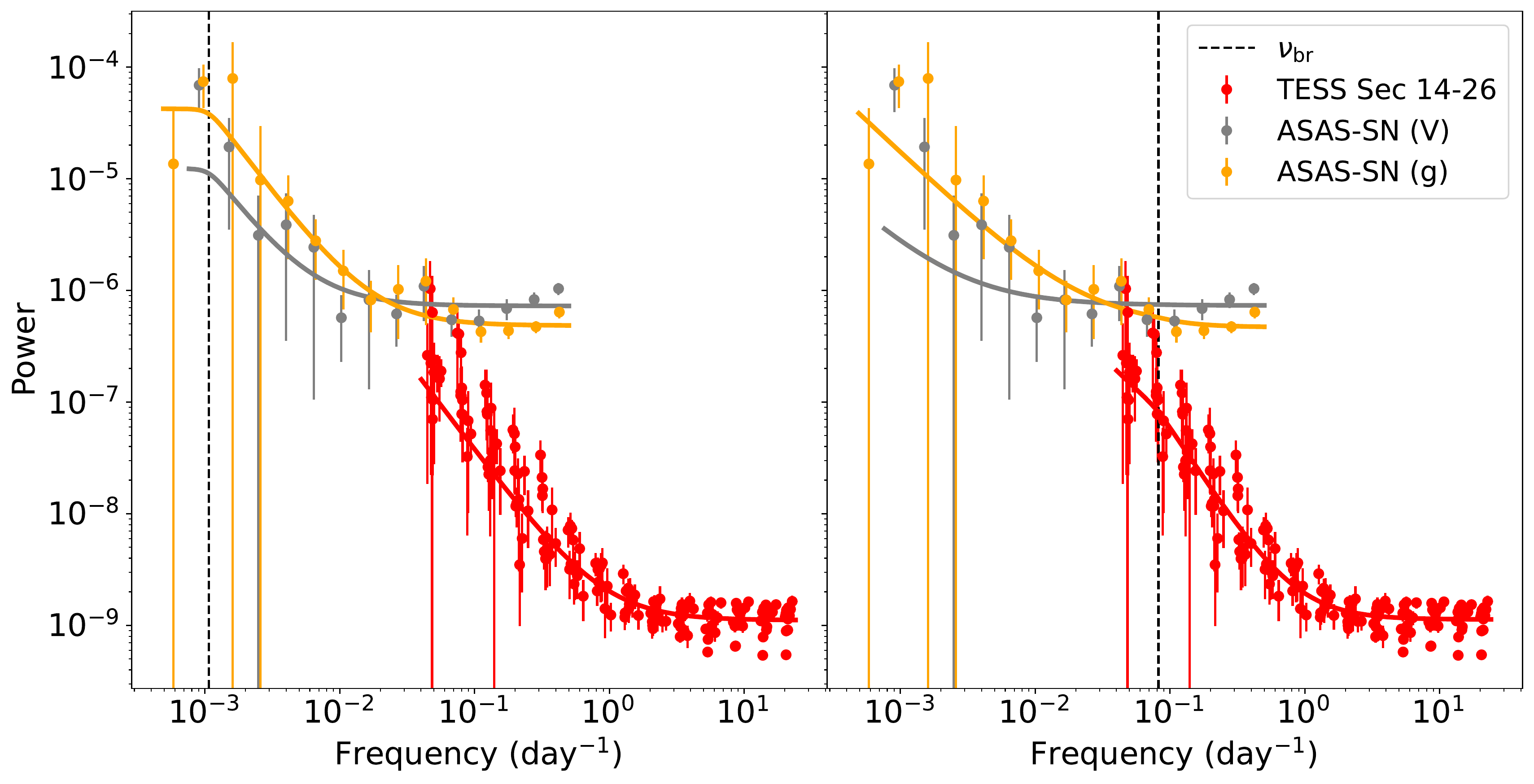}
    \caption{The PSDs with the broken power law fit of PG 1613+658. The PSDs in red labelled TESS sectors 14-26 are the PSDs of the individual sectors.}
    \label{pg1613+658psdfit}
\end{figure}

We constructed the PSD for the combined TESS light curve using the same procedures to examine if the low-frequency regime of the combined TESS PSDs is consistent with those measured by ASAS-SN, up to a normalisation difference. To do so, we estimated and subtracted the white noise of each PSD. We compare the PSDs of the continuous TESS light curve, the PSDs of ASAS-SN V-band and g-band light curves, and the average of the 13 PSDs for the individual TESS sectors in Figure \ref{pg1613+658psd}, finding that the TESS and ASAS-SN PSDs are consistent in the overlapping regions and have similar slopes at low frequencies. We did not use the PSD from all 13 sectors because it has strong artificial peaks at $14n$ ($n=1,2,3,...$) days created by the observing cadence that strongly affects the fits to the PSD. The results for the broken power law PSD fits are shown in Figure \ref{pg1613+658psdfit}. Break frequencies measured are $\log(\nu_{\textrm{br,hig}}/\textrm{day}^{-1})=-1.09\pm0.03$ and $\log(\nu_{\textrm{br,low}}/\textrm{day}^{-1})=-2.97\pm0.01$, which are within the typical range of break frequencies of the AGN sample.

\section{Results}
\label{secresults}

\begin{figure}
    \centering
    \includegraphics[width=\hsize]{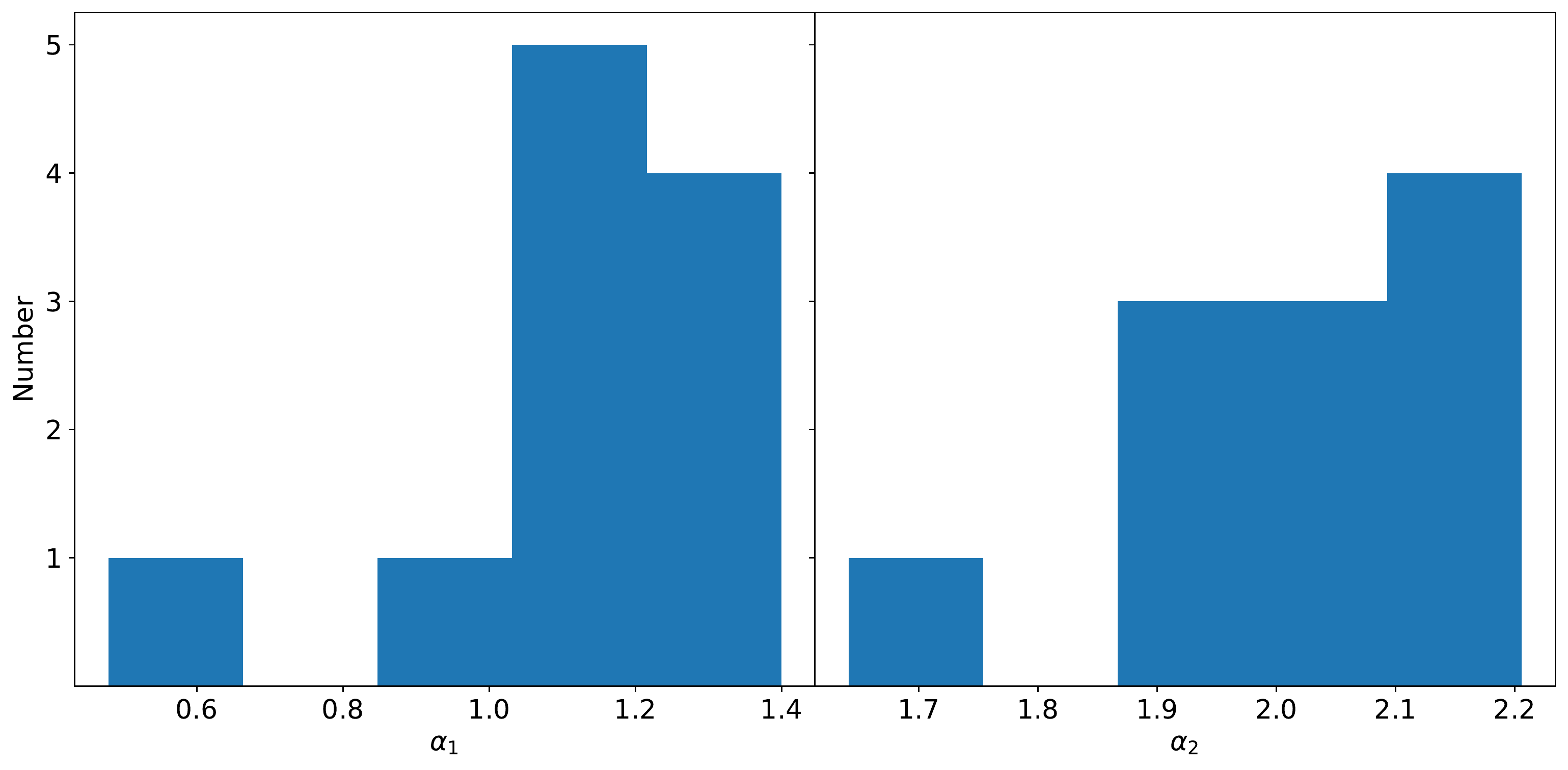}
    \caption{Distribution of the low-frequency (left, $\alpha_1$) and high-frequency (right, $\alpha_2$) slopes for the broken power law fits using $\nu_{\textrm{br,hig}}$.}
    \label{slopes}
\end{figure}

From the broken power law fits, we find two sets of break frequencies, low- and high-frequency breaks with the dividing line at around $10^{-2}$ day$^{-1}$. The low frequency breaks $\nu_{\rm{br, low}}$ are measured in objects with only ASAS-SN data, joint fits with TESS and ASAS-SN with a single $\chi^2_{\textrm{red}}$ minimum below $10^{-2}$ day$^{-1}$, as well as the low frequency $\chi^2_\textrm{red}$ minimum for joint fits with two $\chi^2_{\textrm{red}}$ minima. These break frequencies and their associated slopes are similar to those from the DRW fits, meaning that the break frequencies are close to DRW break frequencies and power law indices are about $\alpha_1=0$ and $\alpha_2=2$. The high frequency breaks $\nu_{\rm{br, hig}}$, meaning a $\chi^2_{\textrm{red}}$ minimum at frequencies higher than $10^{-2} \textrm{ day}^{-1}$, are prominent in most targets with both ASAS-SN and TESS data. These $\nu_{\rm{br, hig}}$ are typically in the overlapping region of the ASAS-SN and TESS power spectra. For ASAS-SN, the PSDs at these frequencies are dominated by white noise. For the fits yielding $\nu_{\rm{br, hig}}$, the power law indices for the low and high frequencies deviate from 0 and 2, especially the low-frequency slopes $\alpha_1$ (Figure \ref{slopes}).

\begin{figure}
    \centering
    \includegraphics[width=\hsize]{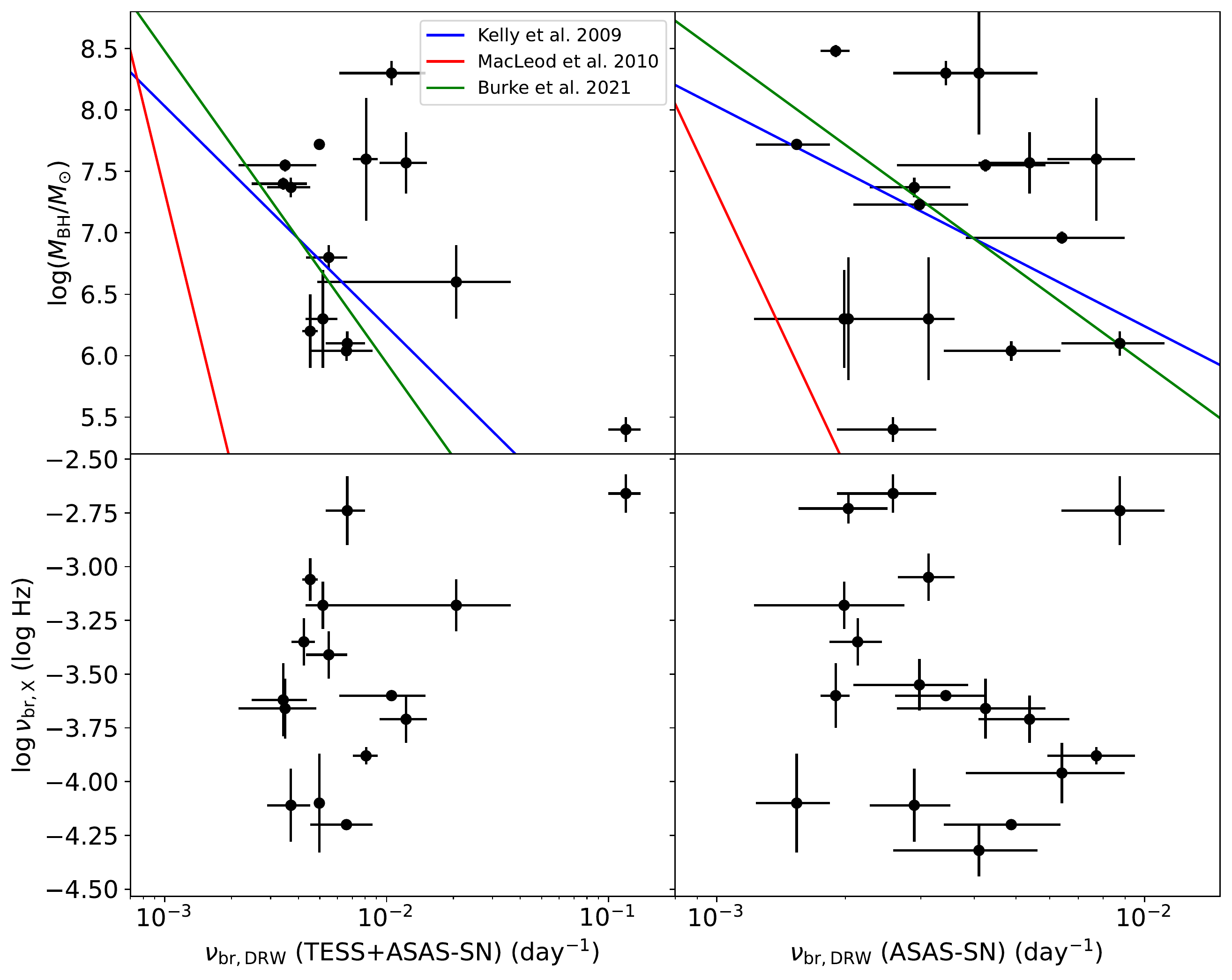}
    \caption{$\nu_{\textrm{br,X}}$ and $M_{\textrm{BH}}$ versus $\nu_{\textrm{br}}=1/(2\pi\tau)$ from DRW model fit on TESS and ASAS-SN PSDs (left), and on ASAS-SN PSDs only (right).}
    \label{breaks_drw}
\end{figure}

\begin{figure}
    \centering
    \includegraphics[width=0.8\hsize]{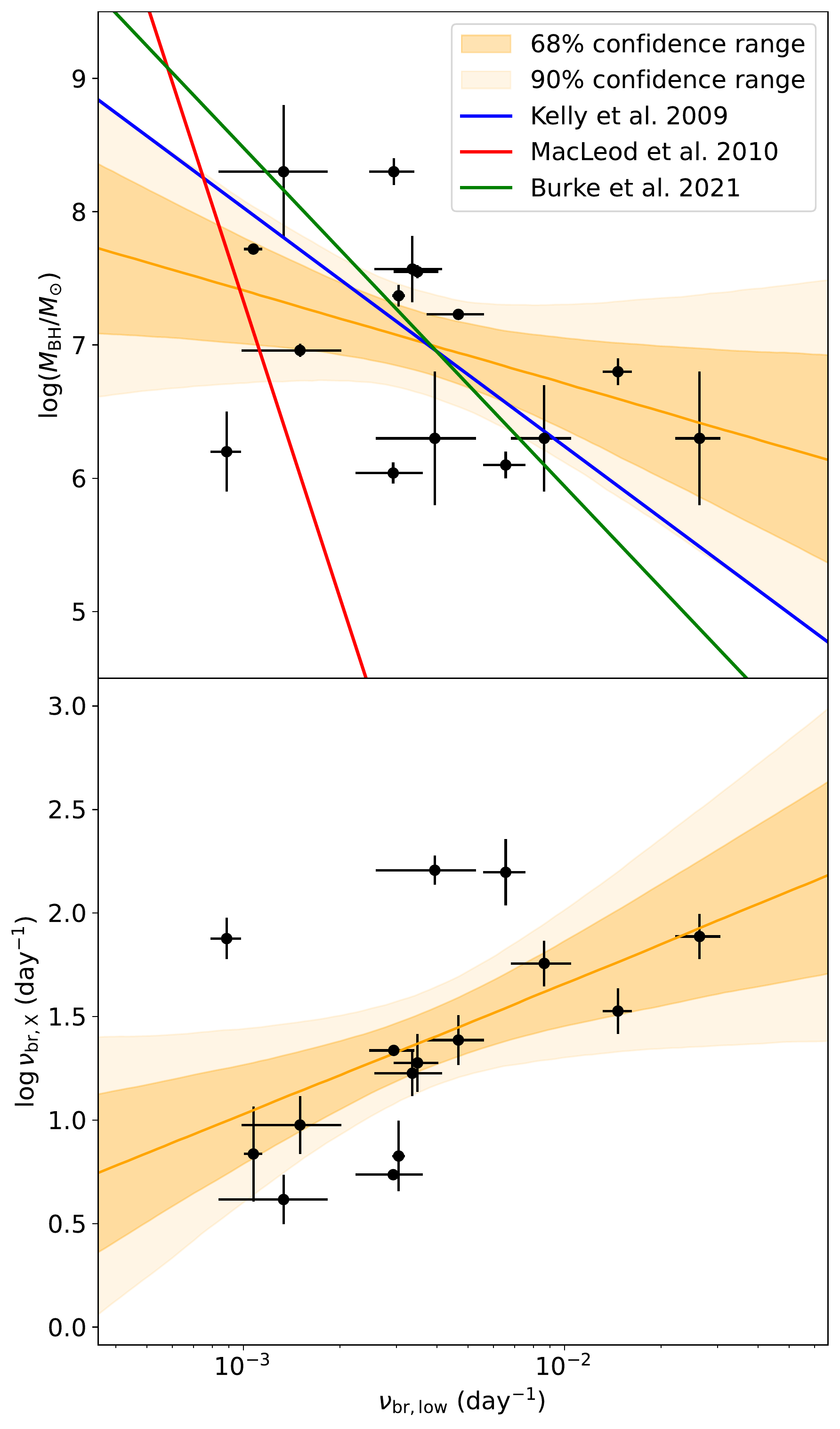}
    \caption{X-ray break frequency $\nu_{\textrm{br,X}}$ and $M_{\textrm{BH}}$ versus low optical break frequency $\nu_{\textrm{br,low}}$ with the best linear fit and its uncertainties. The upper panel shows similar fits from \citet{kelly09}, \citet{macleod10}, and \citet{burke21}.}
    \label{breaks_bpl}
\end{figure}

\begin{figure}
    \centering
    \includegraphics[width=0.8\hsize]{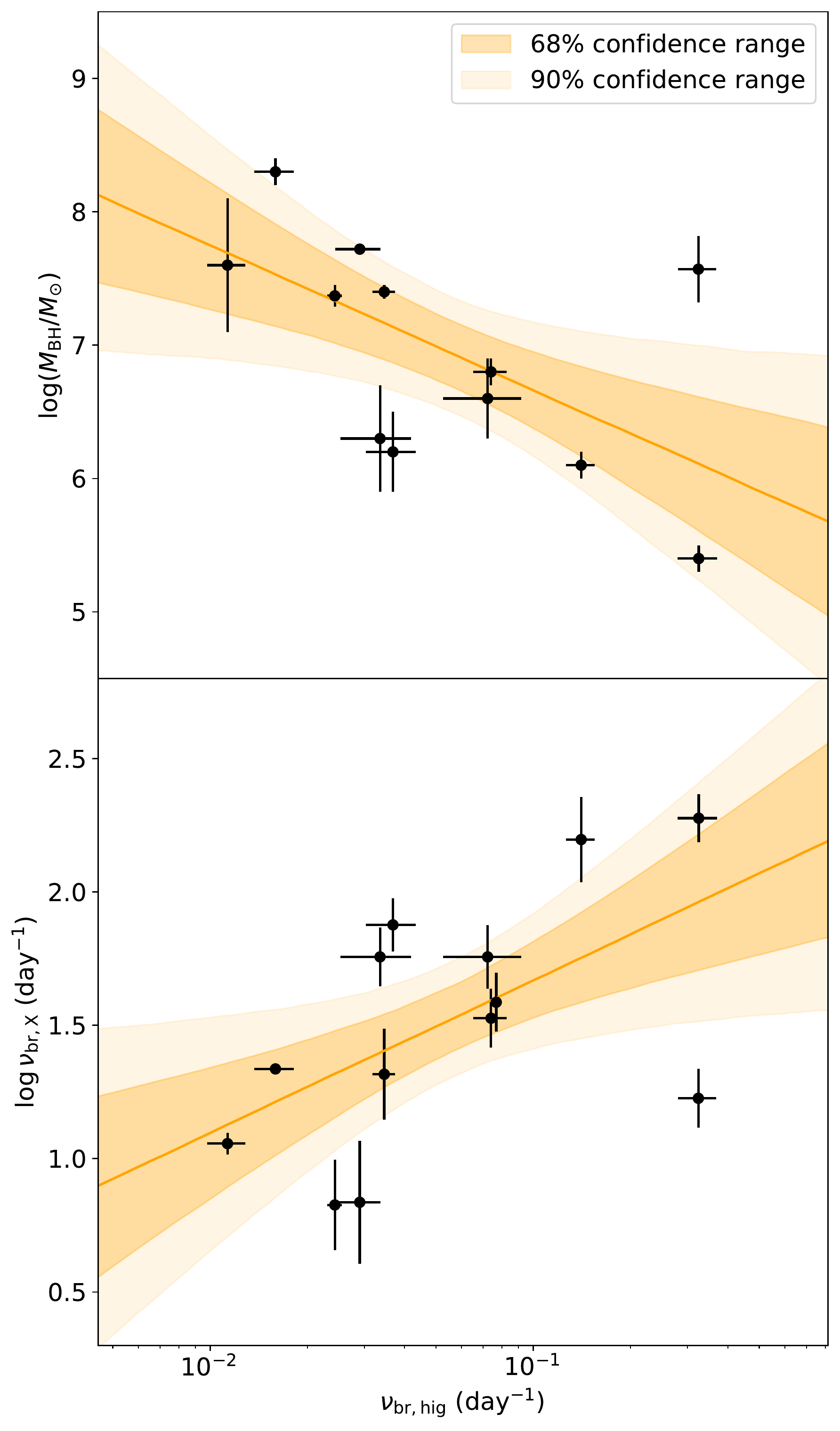}
    \caption{Black hole mass and X-ray break frequency versus the high optical break frequencies along with linear fits and their uncertainties.}
    \label{linfit}
\end{figure}

We have two estimates of the low frequency break from the DRW models (PSD models for either the ASAS-SN or the ASAS-SN plus TESS data), and a set of low and high frequency breaks from the broken power law fits to the ASAS-SN plus TESS PSDs. All breaks related to the DRW models are in the low frequency regime. We compare these breaks with the X-ray break frequencies and black hole masses from \citet{gonzalezmartin18} in Figures~\ref{breaks_drw}, \ref{breaks_bpl}, and \ref{linfit}.
We use the average and uncertainties of X-ray break frequencies presented by \citet{gonzalezmartin18} Figure 3.
To test the relationship between the optical break frequencies and black hole mass and X-ray break frequencies, we computed Pearson's correlation coefficient (Table \ref{correlationtable}). We discuss the low and high frequency regimes separately. 

\begin{table}
\caption{Pearson correlation coefficients between break frequencies and previously measured parameters.}
\label{correlationtable}
\centering
\begin{tabular}{l c c c c}
\hline \hline
 & \multicolumn{2}{c}{$M_{\textrm{BH}}$} & \multicolumn{2}{c}{$\nu_{\textrm{br,X}}$} \\
\cline{2-3} \cline{4-5}
$\nu_{\textrm{br}}$ from & $r$ & $p$ & $r$ & $p$ \\
\hline
BPL fit with  $\nu_{\textrm{br,low}}$   & -0.39 & 0.15 & 0.50 & 0.06 \\
BPL fit with  $\nu_{\textrm{br,hig}}$  & -0.56 & 0.06 & 0.57 & 0.04 \\
BPL fit with corrected $\nu_{\textrm{br,hig}}$  & -0.67 & 0.02 & 0.65 & 0.02 \\
DRW fit with all data                  & -0.43 & 0.13 & 0.48 & 0.07 \\
DRW fit with ASAS-SN data              & -0.06 & 0.82 & -0.16 & 0.55 \\
\hline
\end{tabular}
\end{table}

\begin{table*}
\caption{Properties of the AGN sample. $\log(M/M_{\odot})$ and $\log\nu_{\textrm{br,X}}$ are from \citet{gonzalezmartin18} and references therein. $\log\nu_{\textrm{br,DRW}}$ are from fits using the ASAS-SN PSDs only.}
\label{summarytable}
\centering
\begin{tabular}{l c c c c c}
\hline \hline
     & & $\log\nu_{\textrm{br,X}}$ & $\log\nu_{\textrm{br,DRW}}$ & $\log\nu_{\textrm{br,low}}$ & $\log\nu_{\textrm{br,hig}}$ \\
Name & $\log(M/M_{\odot})$ & (Hz) & (day$^{-1}$) & (day$^{-1}$) & (day$^{-1}$) \\
\hline
MRK 335      & $7.23\pm0.04$ & [-3.9,-2.6] & $-2.53\pm0.13$ & $-2.33\pm0.09$ & \nodata        \\
ESO 113-G010 & \nodata       & [-3.7,-2.9] & $-2.67\pm0.06$ & \nodata        & $-1.11\pm0.02$ \\
Fairall 9    & $8.3\pm0.1$   & -3.6        & $-2.46\pm0.10$ & $-2.53\pm0.07$ & $-1.80\pm0.06$ \\
PKS 0558-504 & $8.48\pm0.05$ & [-4.0,-2.8] & $-2.72\pm0.03$ & \nodata        & \nodata        \\
1H0707-495   & $6.3\pm0.5$   & [-3.5,-2.7] & $-2.51\pm0.07$ & $-1.58\pm0.07$ & \nodata        \\
ESO 434-G040 & $7.57\pm0.25$ & [-4.1,-2.9] & $-2.27\pm0.10$ & $-2.47\pm0.10$ & $-0.49\pm0.06$ \\
NGC 3227     & $6.8\pm0.1$   & [-3.6,-2.5] & \nodata        & $-1.83\pm0.05$ & $-1.13\pm0.05$ \\
REJ 1034+396 & $6.6\pm0.3$   & [-3.5,-2.9] & \nodata        & \nodata        & $-1.14\pm0.12$ \\
NGC 3516     & $7.40\pm0.05$ & [-4.5,-2.8] & \nodata        & \nodata        & $-1.46\pm0.04$ \\
NGC 3783     & $7.37\pm0.08$ & [-4.4,-3.4] & $-2.54\pm0.09$ & $-2.52\pm0.02$ & $-1.61\pm0.02$ \\
NGC 4051     & $6.1\pm0.1$   & [-3.6,-2.2] & $-2.06\pm0.12$ & $-2.18\pm0.07$ & $-0.85\pm0.04$ \\
NGC 4151     & $7.55\pm0.05$ & [-4.0,-2.9] & $-2.37\pm0.16$ & $-2.46\pm0.07$ & \nodata        \\
MRK 766      & $6.2\pm0.3$   & [-3.7,-2.4] & \nodata        & $-3.05\pm0.05$ & $-1.43\pm0.08$ \\
NGC 4395     & $5.4\pm0.1$   & [-3.2,-2.4] & $-2.59\pm0.11$ & \nodata        & $-0.49\pm0.06$ \\
MCG-06-30-15 & $6.3\pm0.4$   & [-3.7,-2.6] & $-2.70\pm0.17$ & $-2.06\pm0.09$ & $-1.47\pm0.11$ \\
IC 4329A     & $8.3\pm0.5$   & [-4.4,-3.0] & $-2.39\pm0.16$ & $-2.88\pm0.16$ & \nodata        \\
Circinus     & $6.04\pm0.08$ & [-4.2,-4.2] & $-2.31\pm0.13$ & $-2.53\pm0.10$ & \nodata        \\
NGC 5506     & $8.1\pm0.2$   & [-3.9,-2.8] & \nodata        & \nodata        & \nodata        \\
NGC 5548     & $7.72\pm0.02$ & [-4.2,-2.9] & $-2.81\pm0.09$ & $-2.97\pm0.03$ & $-1.54\pm0.07$ \\
NGC 6860     & $7.6\pm0.5$   & [-4.0,-2.9] & $-2.11\pm0.10$ & \nodata        & $-1.95\pm0.06$ \\
ARK 564      & $6.3\pm0.5$   & [-3.1,-2.2] & $-2.69\pm0.10$ & \nodata        & \nodata        \\
NGC 7469     & $6.96\pm0.05$ & [-4.1,-2.7] & $-2.19\pm0.18$ & $-2.82\pm0.15$ & \nodata        \\
\hline
\end{tabular}
\end{table*}

\subsection{Low-frequency optical breaks}
From the three sets of low-frequency optical breaks, we focus on those from the broken power law fits, since they correlate better with the black hole mass and X-ray breaks of the sample. 

The ranked correlation test shows that the low frequency break from the broken power law fits weakly correlate with both the black hole mass and the X-ray break frequencies with null probabilities of 0.15 and 0.06, respectively. The break versus black hole mass relation is quite consistent with previous relations (Figure~\ref{breaks_bpl}).
We fit the relationships using

\begin{equation}
    \log(\frac{\nu_{\textrm{br,X}}}{\textrm{day}^{-1}}) = 
    \beta_1 \log(\frac{\nu_{\textrm{br,low}}}{10^{-1}\textrm{day}^{-1}}) + \delta_1
\end{equation}
and

\begin{equation}
    \log(\frac{M_{\textrm{BH}}}{M_{\odot}}) = 
    \beta_2 \log(\frac{\nu_{\textrm{br,low}}}{10^{-1}\textrm{day}^{-1}}) + \delta_2.
\label{mbh-nubrlow}
\end{equation}
We find that $\beta_1=0.63^{+0.35}_{-0.36}$, $\delta_1=2.30^{+0.51}_{-0.54}$, $\beta_2=-0.69^{+0.59}_{-0.60}$, and $\delta_2=6.01^{+0.89}_{-0.87}$, with the results shown in Figure \ref{breaks_bpl}. 
Although the relation between $\nu_{\textrm{br,low}}$ and $M_{\textrm{BH}}$ is not statistically significant in this paper, we list our relation as a comparison with previous studies, who found the relation significant \citep[e.g.,][]{kelly09, macleod10, burke21}.
PG 1613+658, the Seyfert I galaxy with 13 continuous TESS sectors discussed earlier, has a reverberation mapping black hole mass measurement of $\log(M_{\textrm{BH}}/M_{\odot})=8.38^{+0.25}_{-0.16}$ \citep{kaspi00}. Using the break frequency from the PSD of the continuous TESS light curve and this $M_{\textrm{BH}}$-$\nu_{\textrm{br}}$ relationship (Equation \ref{mbh-nubrlow}), we get $\log(M_{\textrm{BH}}/M_{\odot})=7.39^{+1.43}_{-1.44}$, which agrees with the previous measurement.

\subsection{High-frequency optical breaks}
The high frequency optical breaks were detected in the broken power law fits to the combined ASAS-SN and TESS PSDs. They are all at frequencies above $10^{-2}$~day$^{-1}$, and resemble the steepening of optical PSDs found in Kepler data (\citealt{mushotzky11}; \citealt{kasliwal15}; \citealt{aranzana18}; \citealt{smith18}). Here, we can compare them with the X-ray break frequencies measured for the same set of objects. The high-frequency optical $\nu_{\textrm{br,hig}}$ and X-ray $\nu_{\textrm{br,X}}$ break frequencies are given in Table \ref{summarytable} along with $M_{\textrm{BH}}$ and shown in Figure \ref{linfit}. The ranked correlation test shows that the high frequency optical breaks and X-ray breaks are correlated with a null probability of 4\%. We also found a hint of a correlation between the high frequency breaks and the black hole masses, but at lower significance with a null probability of 6\%. Figure \ref{linfit} shows fits to the correlations using

\begin{equation}
    \log(\frac{\nu_{\textrm{br,X}}}{\textrm{day}^{-1}}) = 
    \beta_3 \log(\frac{\nu_{\textrm{br,hig}}}{10^{-1}\textrm{day}^{-1}}) + \delta_3
\end{equation}
and

\begin{equation}
    \log(\frac{M_{\textrm{BH}}}{M_{\odot}}) = 
    \beta_4 \log(\frac{\nu_{\textrm{br,hig}}}{10^{-1}\textrm{day}^{-1}}) + \delta_4.
\label{mbh-nubr}
\end{equation}
to find that $\beta_3=0.57^{+0.30}_{-0.29}$, $\delta_3=1.67^{+0.15}_{-0.14}$, $\beta_4=-1.09^{+0.56}_{-0.56}$, and $\delta_4=6.66^{+0.29}_{-0.29}$.
If we apply this $M_{\textrm{BH}}$-$\nu_{\textrm{br}}$ relationship to PG 1613+658, we get $\log(M_{\textrm{BH}}/M_{\odot})=6.76^{+0.30}_{-0.30}$, which does not agree with the previous measurement. Further studies are needed to evaluate the validity of this relationship.

We find that the slopes of the broken power law fits do not have a strong correlation with the black hole mass or the break frequency. The Pearson's correlation test parameters are $r=-0.17$ and $p=0.63$ for $\alpha_1-M_{\textrm{BH}}$, $r=0.33$ and $p=0.32$ for $\alpha_1-\nu_{\textrm{br,hig}}$, $r=-0.38$ and $p=0.28$ for $\alpha_2-M_{\textrm{BH}}$, and $r=-0.22$ and $p=0.51$ for $\alpha_2-\nu_{\textrm{br,hig}}$ pairs.

\subsection{Testing the PSD fitting method}
To determine whether our PSD fitting method is robust and ensure that the high frequency breaks are not simply artefacts of combining two sets of PSDs, we selected a number of model parameters, simulated light curves, and performed the PSD fits method as described in Section~\ref{secmethods} to test if the true model parameters can be recovered. We used the method of \citet{timmer95} to simulate light curves of a given PSD. Then we added white noise and applied the same sampling windows as the observations. To account for potential misestimation of the measurement noise, we used the three-point method to scale the original measurement uncertainty; we checked three consecutive data points within a certain time frame, find the linear fit, and find the scaling factor for the measurement uncertainties such that $\chi^2=1$. This process was repeated for each ASAS-SN/TESS light curve.
For this testing, we started with a specific set of parameters, and changed one parameter at a time. We used two different sets of initial parameter sets: $\log(\nu_{\textrm{br}}/\textrm{day}^{-1})=-2.5$, $\alpha_1=0.3$, and $\alpha_2=2.1$, which correspond to the typical values for low-frequency breaks, and $\log(\nu_{\textrm{br}}/\textrm{day}^{-1})=-1.5$, $\alpha_1=1.2$, and $\alpha_2=2.1$, which correspond to the typical values for high-frequency breaks. 
The ranges of parameters are $0.0\le\alpha_1\le1.5$, $1.5\le\alpha_2\le3.0$ with steps of 0.3, and $-2.8\le\log\nu_{\textrm{br}}\le-1.0$ with steps of a 0.3 dex.

\begin{figure*}
    \centering
    \includegraphics[width=0.8\hsize]{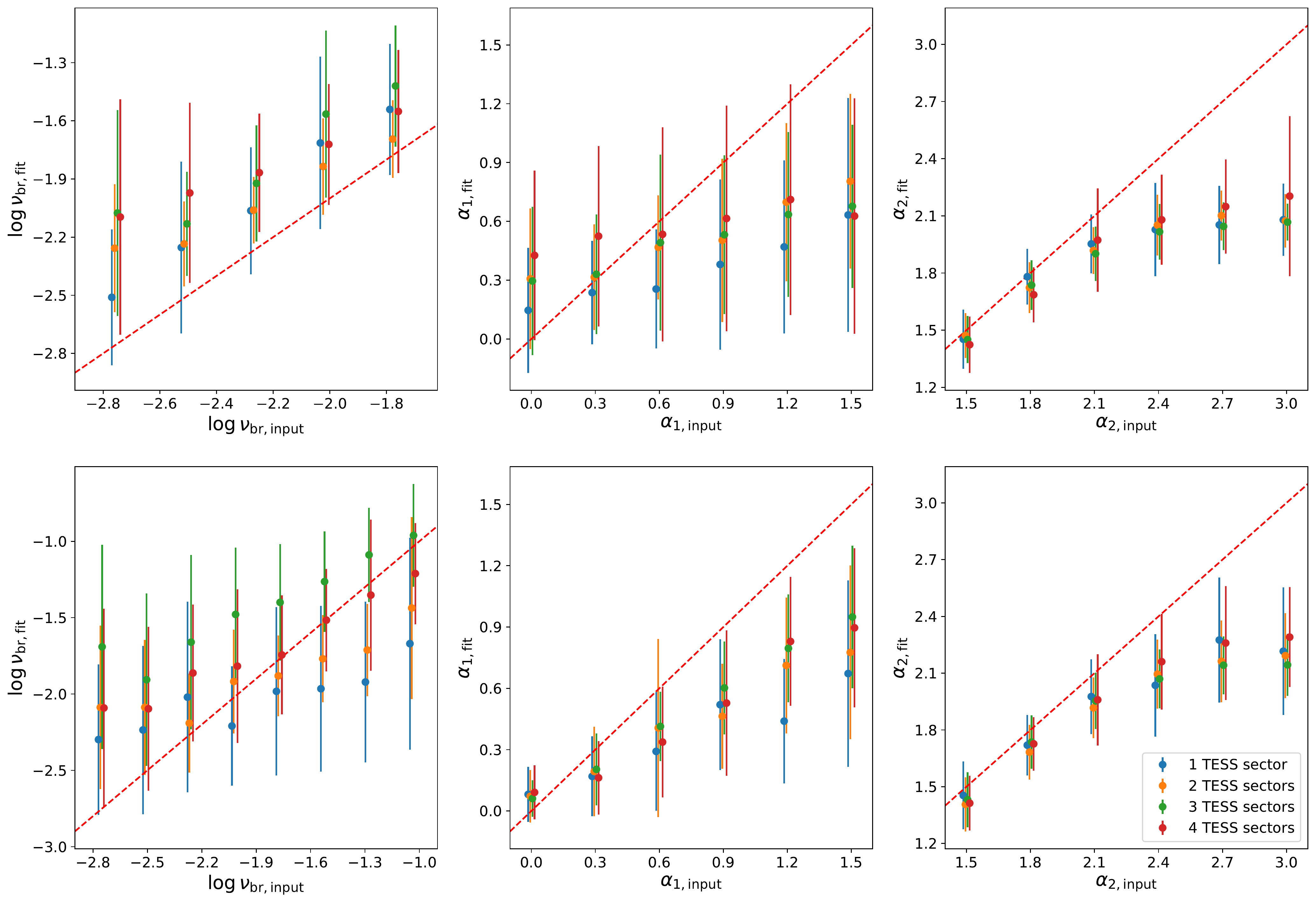}
    \caption{Comparisons of BPL model parameters from the PSD fitting method and the true values used for light curve simulations. Top: variations from the low-frequency break parameters. Bottom: variations from the high-frequency break parameters. The red dashed line in each panel represents the one-to-one relation. The data points are slightly offset along the x-axis to make them distinguishable.}
    \label{sim_to_true}
\end{figure*}

The results are shown on Figure \ref{sim_to_true}. We observe varying levels of biases between the best-fit and input model parameter values. However, the biases can be estimated.  For the break frequency, we find that the input and measured break frequencies have an approximately linear relationship, but with offsets. With a shallow $\alpha_1$ (varying from the low-frequency break setup), measured break frequencies deviate more towards larger values at lower frequencies. With a steeper $\alpha_1$ (varying form the high-frequency break setup), cases with more TESS sectors show better agreement, while having fewer TESS sectors appears to lead to almost constant values. For both cases, the discrepancy is greatest at the lowest break frequency, but that may be caused by the fact that it is near the PSD lower frequency measurement limit. For power law indices, it appears the measured values have some upper limit to what this method can measure.

\begin{figure}
    \centering
    \includegraphics[width=\hsize]{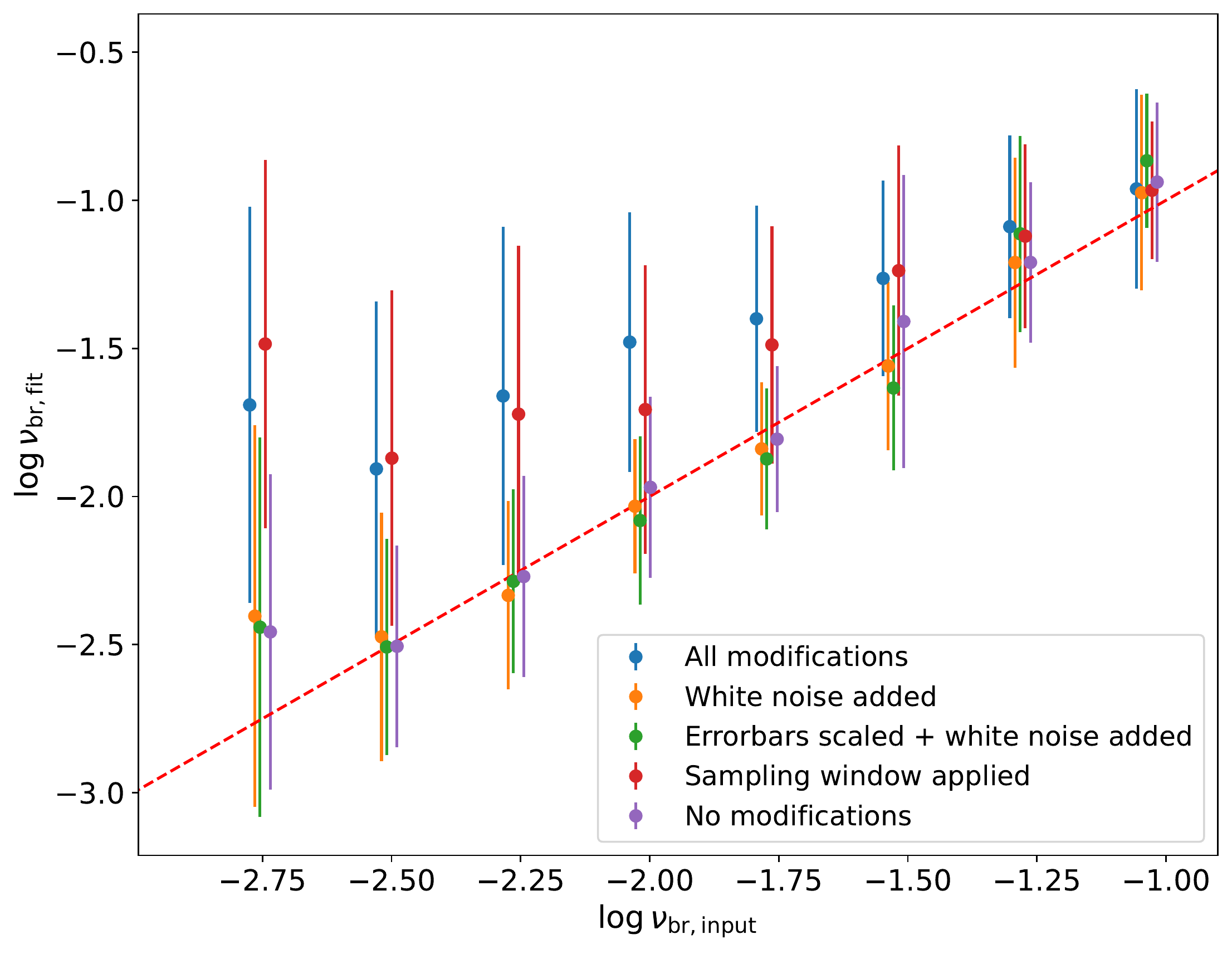}
    \caption{Measured break frequencies with different levels of modifications on the light curves versus the input break frequencies for the light curve simulations. The red dashed line represents the one-to-one relation. The data points are slightly offset along the x-axis to make them distinguishable.}
    \label{compare_steps}
\end{figure}

To determine which factor contributes the most to the observed discrepancies, we tried using simulated light curves by adding different observational biases one at a time. We tested adding white noise, our treatment of measurement uncertainties, and the effect of observing sampling windows. We used the three simulated TESS sectors for the tests. We found that the light curve sampling window affects the results the most (Figure \ref{compare_steps}), while the white noise and our treatment of measurement uncertainties had little effect on the measured PSD parameters. Since the sampling window effect is significant for ASAS-SN, it makes sense that the discrepancy is greater when there are fewer TESS sectors, where the ASAS-SN PSDs contribute more to the fit.
In summary, figures \ref{sim_to_true} and \ref{compare_steps} show that combining two sets of PSDs from TESS and ASAS-SN introduces a bias on measured break frequency towards higher values. However, the measured values are mostly consistent with true values, suggesting that the $\nu_{\textrm{br,hig}}$ we measured are not due to combining PSDs.

\begin{figure}
    \centering
    \includegraphics[width=0.8\hsize]{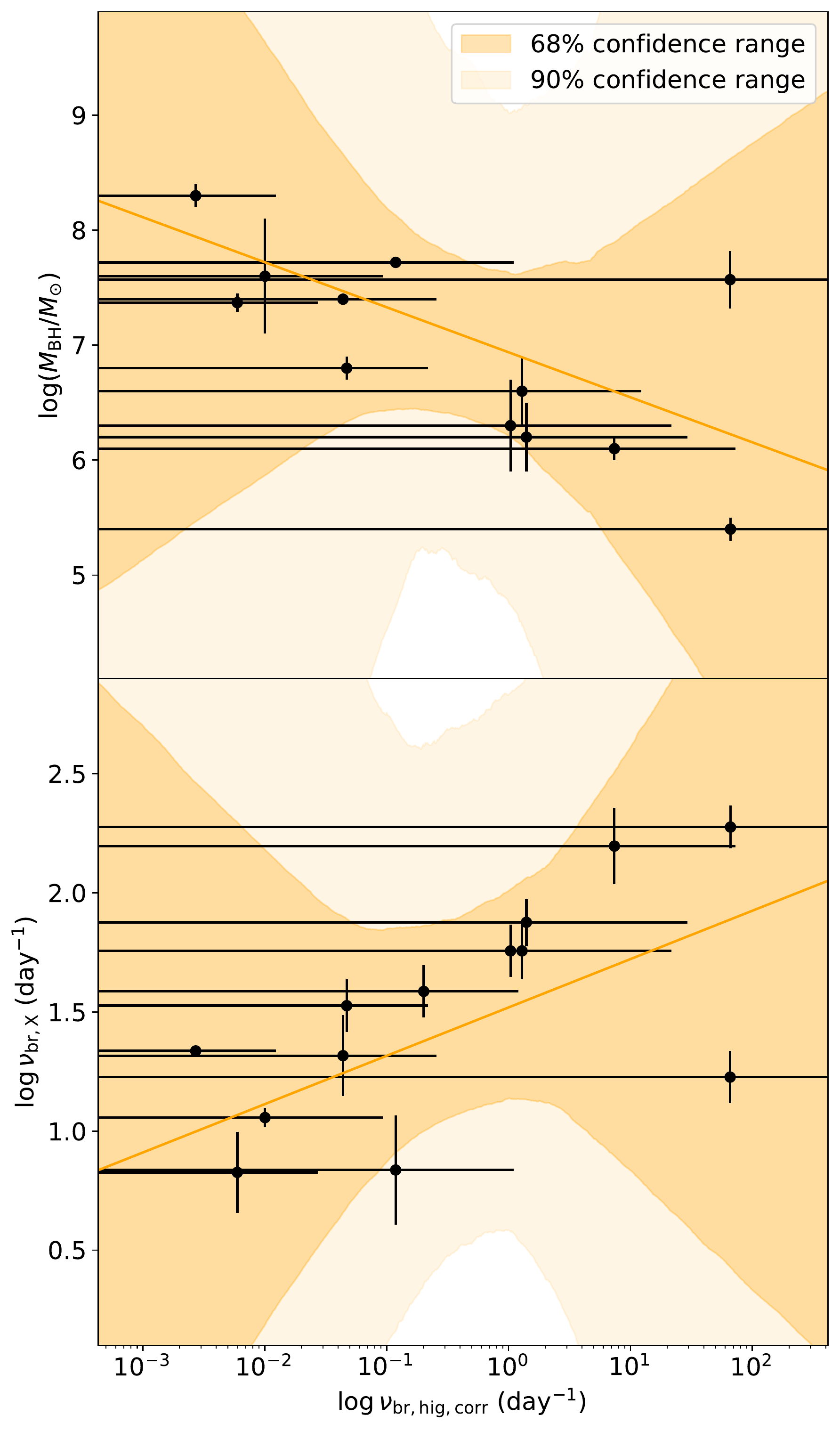}
    \caption{Same as Figure \ref{linfit}, but with $\nu_{\textrm{br,hig}}$ corrected for distortion.}
    \label{linfit_corr}
\end{figure}

We also attempted calculating the correction for the measured high frequency break. After correction, the linear trend is still there, but the errorbars are much larger and the slope is much shallower (Figure \ref{linfit_corr}). The newly measured correlation parameters are $\beta_3=0.20^{+0.61}_{-0.69}$, $\delta_3=1.52^{+0.49}_{-0.38}$, $\beta_4=-0.39^{+1.13}_{-1.11}$, and $\delta_4=6.94^{+0.70}_{-0.73}$, with the correlation null probabilities of 0.02 and 0.02.
With the break frequency correction, $\log(\nu_{\textrm{br,hig}}/\textrm{day}^{-1})$ of PG 1613+658 becomes $-0.68\pm1.72$, and with the new correlation parameters, its black hole mass is $\log(M_{\textrm{BH}}/M_{\odot})=6.76^{+1.58}_{-1.53}$. Compared to the reverberation mapping measurement, the difference in central values is still large, the two values agree within $2\sigma$ because of the large uncertainties in the correlation.

\subsection{Multi-variable correlation plane}

A number of studies have explored the correlation between the X-ray break frequency and black hole mass, and also examines how other parameters, such as bolometric luminosity and obscuration, can better constrain that correlation \citep[e.g.,][]{mchardy06, gonzalezmartin12, gonzalezmartin18}. We also tested if our newly found correlations are affected by the obscuration, bolometric luminosity, X-ray luminosity, optical luminosity, or optical-to-X-ray luminosity ratio. \citet{gonzalezmartin18} reports the range of values of $N_H$ and $L_{\textrm{bol}}$ for the sample. The reported $L_{\textrm{bol}}$ is derived from the X-ray luminosity at 2--10 keV:
\begin{equation}
    \log\bigg(\frac{L_{\textrm{bol}}}{L(2-10\textrm{ keV})}\bigg)=1.54+0.24\mathcal{L}+0.012\mathcal{L}^2-0.0015\mathcal{L}^3,
\end{equation}
where $\mathcal{L}=\log(L_{\textrm{bol}}/L_{\odot})-12$ \citep{marconi04}. We use the middle values from the range of $N_H$ and $L_{\textrm{bol}}$ reported by \citet{gonzalezmartin18} and derive $L_{\textrm{X}}$ from $L_{\textrm{bol}}$. For optical luminosity, we use the median ASAS-SN $V$-band magnitude, after correcting for galactic reddening and K-correction, to estimate the $V$-band luminosity of the sample.

We fit the multi-variable correlation,
\begin{equation}
    Y = \beta \log\nu_{\textrm{br,opt}} + \gamma X + \delta,
    \label{multieq}
\end{equation}
where $Y$ is either the $\log\nu_{\textrm{br,X}}$ or $\log M_{\textrm{BH}}$ and $\log\nu_{\textrm{br,opt}}$ is the optical break frequency ($\log\nu_{\textrm{br,low}}$, $\log\nu_{\textrm{br,hig}}$, or $\log\nu_{\textrm{br,hig,corr}}$), $X$ is either $\log N_H$, $\log L_{\textrm{bol}}$, $\log L_{\textrm{X}}$, $\log L_V$, or $\log (L_V/L_{\textrm{X}})$ and $\beta$, $\gamma$, and $\delta$ are constants for fitting. We tested all combinations with $X$, $Y$, and $\log\nu_{\textrm{br,opt}}$. 
In most cases, the best-fit $\beta$ values are consistent with the $1\sigma$ range of the original two-parameter correlation values.
For some of the combinations, $|\beta|$ approaches zero for the best fit, and the multi-variable correlation is driven by the parameters $Y$ and $X$.  These cases are not related to the main goal of this paper, which is to explore the correlations related to $\nu_{\textrm{br,opt}}$ of either high or low frequencies.  Thus, we fixed $\beta$ values at the best-fit values from the two-parameter correlations for these cases.

\begin{figure*}
    \centering
    \includegraphics[width=\hsize]{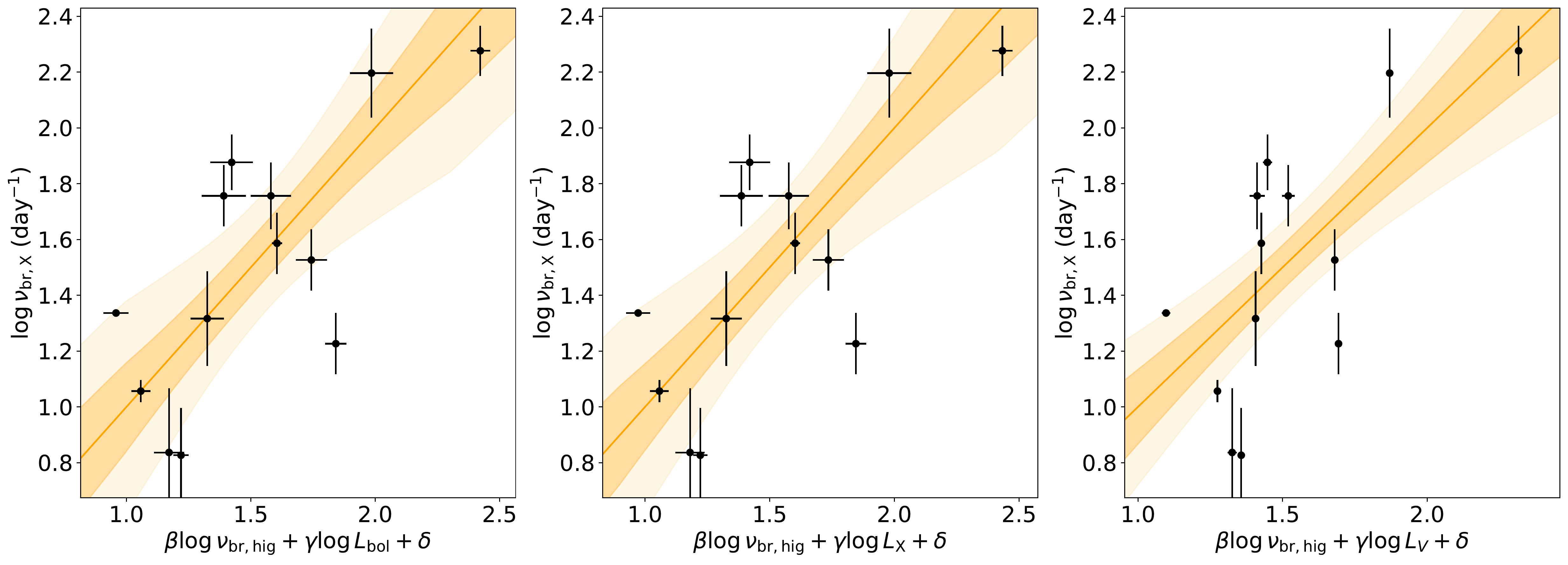}
    \caption{The results of fitting the correlation between $\log\nu_{\textrm{br,X}}$ and $\log\nu_{\textrm{br,hig}}$ with the addition of $L_{\textrm{bol}}$ (left), $\log L_{\textrm{X}}$ (middle), or $\log L_V$ (right). The dark and lightly shaded regions represent the 68\% and 95\% confidence ranges, respectively.}
    \label{multivariate_fitting_X_hig}
\end{figure*}

We find that, for all combinations, the correlations improve significantly with the addition of X-ray or optical luminosity and slightly with the addition of optical-to-X-ray luminosity ratio or obscuration. The only exception is $\nu_{\textrm{br,low}}-\log M_{\textrm{BH}}$ correlation with the addition of $\log N_{\textrm{H}}$, which had nearly no improvement.
For all the multi-variate correlations with the additional X-ray or optical luminosity parameter we tested, the median significance has improved to $p=0.01$ compared to the median significance of $p=0.05$ when considering only the original two parameter correlations.
By the nature of bolometric luminosity being derived from the X-ray luminosity, the effect of adding $L_{\textrm{bol}}$ is nearly the same as adding $L_{\textrm{X}}$. For example, the original two-parameter correlation between $\log\nu_{\textrm{br,hig}}$ and $\log\nu_{\textrm{X}}$ has the correlation coefficient of $r=0.57$ with $p=0.04$. With the addition of $\log L_{\textrm{bol}}$, $\log L_{\textrm{X}}$, or $\log L_V$, the correlation coefficient becomes $r=0.73$ with $p=0.005$, $r=0.73$ with $p=0.005$, and $r=0.69$ with $p=0.009$, respectively (Figure \ref{multivariate_fitting_X_hig}). We present the entirety of our multi-variable fit results in Appendix \ref{secmulti}.

\section{Summary and discussion}
\label{secdiscussion}
We study AGN variability over six orders of magnitude from minutes to decade time-scales by combining TESS and ASAS-SN light curves. Our analysis shows that the ASAS-SN and TESS PSDs can be jointly modelled with a normalisation offset due to differences in noise levels and filters. We analysed a bright target PG 1613+658 in the TESS continuous monitoring regions, giving it a TESS light curve spanning more than a year, so that there is a significant overlap between PSDs, and found consistent light curves and PSD measurements.

We also compare our results to scaling relations between the low frequency break and black hole mass from \citet{kelly09}, \citet{macleod10}, and \citet{burke21}. To compare with the multi-variate correlation \citet{macleod10}, we use their relations assuming a rest frame wavelength of  $\lambda_{\textrm{RF}}=8000 \textrm{\r{A}}$, an absolute magnitude of $M_i=-21.3$, and a redshift of $z=0.02$. All three relations are shown in Figures \ref{breaks_drw} and \ref{breaks_bpl}.

The $\nu_{\textrm{br}}-M_{\textrm{BH}}$ relationship found using the various DRW fits do not agree with previous studies very well, though the DRW PSD fits using both TESS and ASAS-SN have a similar trend (Figure \ref{breaks_drw}). Although the correlations between these DRW break frequencies and X-ray break frequencies are not significant, as can be seen in the bottom panels in Figure \ref{breaks_drw}, the X-ray break dependence on the DRW breaks is almost flat, which is contrary to the slope of $\beta_1=0.63^{+0.35}_{-0.36}$ between the low frequency breaks from the broken power law model and X-ray breaks.

\citet{burke20} performed a similar break frequency analysis on NGC 4395. They used the TESS sector 22 light curve to construct the PSD and fit a broken power law. They found a break frequency of $\log(\nu_{\textrm{br}}/\textrm{day}^{-1})=-0.94\pm0.25$, which differs from our value of $\log(\nu_{\textrm{br}}/\textrm{day}^{-1})=-0.49\pm0.06$. This may be due to using both TESS and ASAS-SN data. Using only the TESS PSDs, we find $\log(\nu_{\textrm{br}}/\textrm{day}^{-1})=-0.93\pm0.07$, which agrees with \citet{burke20}.

AGN variability PSDs are generally characterised by power laws with breaks, where distinct break frequencies have been observed in both optical (e.g., \citealt{kelly09}; \citealt{macleod10}; \citealt{simm16}; \citealt{burke21}) and X-ray data (e.g., \citealt{uttley02}; \citealt{markowitz03}; \citealt{gonzalezmartin12}; \citealt{gonzalezmartin18}). Both the optical and X-ray break frequencies are correlated with the black hole mass, but with quite different scaling relations (e.g., \citealt{simm16}; \citealt{burke21}).
The fact that the break frequency and black hole mass have a statistically significant correlation suggests that the AGN variability is closely related to accretion physics. 

This paper presents the first analysis comparing optical and X-ray PSDs by utilizing the all-sky nature of ASAS-SN and TESS surveys. We find a set of high frequency breaks ($\gtrsim 10^{-2}$day$^{-1}$) in the joint optical light curves. These high frequency breaks were previously observed as a steepening of the optical PSD at high frequencies (\citealt{mushotzky11}; \citealt{kasliwal15}; \citealt{aranzana18}; \citealt{smith18}). We find a potential correlation between the X-ray and high-frequency optical break frequencies. 
The slope of the power law relation of $\beta_3 = 0.59^{+0.27}_{-0.27}$ is close to a relation of $\beta_3=0.5$, given the uncertainty, shallower than a 
simple proportional relation of $\beta_3=1$. We can interpret the offset between the optical and X-ray break frequencies as the light crossing time between the X-ray and optical emission regions. A slope shallower than a proportional relation implies that the X-ray emission region is relatively smaller than the optical for greater $M_{\textrm{BH}}$.
There is also an indication that the high frequency optical breaks may be correlated with the black hole mass, though weakly. This could be a new scaling relationship $\nu_{\textrm{br}}-M_{\textrm{BH}}$, different from those found in previous studies, providing an alternative method to estimate the black hole mass.
However, with corrections for sampling, the correlations between the high frequency breaks and X-ray break frequencies and black hole mass become less significant. A larger sample is needed to better establish the correlation with more accurate slope measurements. 

We further explore whether the correlations discussed in this paper are affected by
the X-ray luminosity, optical luminosity, optical-to-X-ray luminosity ratio, or obscuration.  We find that including either the X-ray, optical, or bolometric luminosity as the third parameter can significantly improve the correlation significances, the optical-to-X-ray luminosity ratio or obscuration modestly or slightly improves the correlation significances.
With the inclusion of an additional parameter, the dependence between the X-ray and high-frequency optical break has a power law slope of $\beta_3$ in the range of 0.1 to 0.8 with a median value of 0.52 (Table~\ref{multitable}), which is consistent with the original two parameter slope of $\beta_3 = 0.59\pm0.27$.

If the optical and X-ray frequencies are indeed correlated, we can speculate on its implications.
The X-ray emission is produced from inverse Compton scattering between UV seed photons from the accretion disc and electrons in the corona. The X-ray emission then irradiates the accretion disc, heating the disc and modifying the optical emission. 
Therefore, the optical and X-ray variability should show correlations. In fact, short timescale correlations between optical and X-ray light curves are well established and used as a tool for continuous reverberation mapping studies to constrain optical and X-ray emission sizes (e.g., \citealt{shappeeprieto14}; \citealt{edelson15}; \citealt{troyer16}; \citealt{mchardy18}). The correlation we find between the X-ray and optical PSD break frequencies provides independent support that the optical and X-ray emissions are closely related on short time scales. 
Since the optical PSD on these scales deviates from the extrapolation from longer time-scales, the variation of the X-ray emission, through ``reprocessing'', can be the main driver for the interaction between optical and X-ray emission at short time-scales.

The X-ray emission, which can vary on very short timescales, is speculated to come from the innermost part of the AGN, near the SMBH. On the other hand, the optical emission is believed to come from the accretion disc.
Studies in quasar microlensing (e.g., \citealt{morgan08}; \citealt{dai10}) and reverberation mapping (e.g., \citealt{shappeeprieto14}; \citealt{edelson15}; \citealt{cackett18}; \citealt{jha22}) find that the sizes of X-ray and optical emission regions range from $10^{14}$ to $10^{15}$ cm and from $10^{15}$ to $10^{16}$ cm, respectively.
This scale is comparable to our observed difference in characteristic timescales, with $\tau=1/(2\pi\nu_{\textrm{br}})\sim10^{-2}$ day for the X-rays and $\sim10^0$ days for the optical.

In the low frequency regime ($\lesssim 10^{-2}$day$^{-1}$), we also found a correlation between the optical and X-ray breaks, with a similar slope $\beta_1 = 0.63^{+0.35}_{-0.36}$.
Our analysis results show that the DRW model fits the ASAS-SN PSDs well.
We also roughly recover the DRW break versus black hole mass relation measured in the literature.
When high-frequency TESS PSDs are introduced, the DRW work less well. 
The broken power law fits show that the power law indices deviate from the DRW values of 0 and $-2$ as shown on the distributions on Figure \ref{slopes}, especially for the low-frequency slope. 
A few objects potentially have two break frequencies from the broken power law fits (Figure \ref{redchisq_2valleys}). 
This means that AGN variability is more complicated and may be governed by different physical processes at different scales. For example, the longer-term variability could be mainly described by a DRW model, physically related to MRI or other disc variability mechanisms (e.g., \citealt{kelly09}; \citealt{macleod10}), while the short-term optical variability is strongly affected by other variability mechanisms, such as the reprocessing of X-rays.

\section{Data availability}
Tables \ref{lctable} and \ref{lctable_tess} are available in electronic form at the CDS via anonymous ftp to cdsarc.u-strasbg.fr (130.79.128.5) or via http://cdsweb.u-strasbg.fr/cgi-bin/qcat?J/A+A/.

\begin{acknowledgements}  We thank the anonymous referee for constructive suggestions.
H.Y. and X.D. would like to acknowledge NASA funds 80NSSC22K0488, 80NSSC23K0379 and NSF fund AAG2307802. H.Y. thank the Avenir fellowship. We also thank the OU Supercomputing Center for Education \& Research (OSCER) at the University of Oklahoma for providing the computing resources.
\end{acknowledgements}

\bibliographystyle{aa}
\bibliography{reference}

\begin{appendix}

\onecolumn

\section{Light curve compilation}
Here, we present the compilation of light curves used for this study in both text and visual formats. ASAS-SN and TESS light curves are shown on Tables \ref{lctable} and \ref{lctable_tess} and Figures \ref{lcs_compilation_1}, \ref{lcs_compilation_2}, and \ref{lcs_compilation_3}.

\begin{table*}[h!]
\caption{The ASAS-SN light curves used for this study. This table is available in a machine-readable form.}
\label{lctable}
\centering
\begin{tabular}{l c c c c}
\hline \hline
Name & Filter & Date (HJD) & Flux (mJy) & Flux err (mJy) \\
\hline
MRK 335      & V & 2456233.79722 & 8.67400 & 0.15044 \\
             & V & 2456461.02849 & 8.27800 & 0.25125 \\
             & V & 2456470.02629 & 8.55800 & 0.38962 \\
             &   & \nodata       & \nodata & \nodata \\
MRK 335      & g & 2458017.75063 & 5.93600 & 0.15929 \\
             & g & 2458035.89937 & 5.95800 & 0.15955 \\             & g & 2458035.89942 & 6.47800 & 0.16570 \\
             &   & \nodata       & \nodata & \nodata \\
ESO 113-G010 & V & 2456789.93382 & 7.86700 & 0.38656 \\
             & V & 2456805.91323 & 7.31200 & 0.10935 \\
             & V & 2456809.90571 & 7.12700 & 0.12181 \\
             &   & \nodata       & \nodata & \nodata \\
\hline
\end{tabular}
\end{table*}

\begin{table*}[h!]
\caption{The TESS light curves used for this study. This table is available in a machine-readable form. $^1$TJD = JD - 2457000.}
\label{lctable_tess}
\centering
\begin{tabular}{l c c c c}
\hline \hline
Name & Sector & Date (TJD)$^1$ & Flux (mJy) & Flux err (mJy) \\
\hline
ESO 113-G010 & 1 & 1325.315786 & 8.58287 & 0.01241 \\
             & 1 & 1325.336620 & 8.58208 & 0.01249 \\
             & 1 & 1325.357454 & 8.61625 & 0.01255 \\
             &   & \nodata     & \nodata & \nodata \\
ESO 113-G010 & 2 & 1354.107720 & 8.64351 & 0.01330 \\
             & 2 & 1354.128553 & 8.60523 & 0.01331 \\
             & 2 & 1354.149386 & 8.61716 & 0.01331 \\
             &   & \nodata       & \nodata & \nodata \\
Fairall 9    & 2 & 1354.107720 & 13.76622 & 0.01327 \\
             & 2 & 1354.128553 & 13.81294 & 0.01326 \\
             & 2 & 1354.149386 & 13.79073 & 0.01326 \\
             &   & \nodata       & \nodata & \nodata \\
\hline
\end{tabular}
\end{table*}

\begin{figure*}[h!]
    \centering
    \includegraphics[width=0.95\hsize]{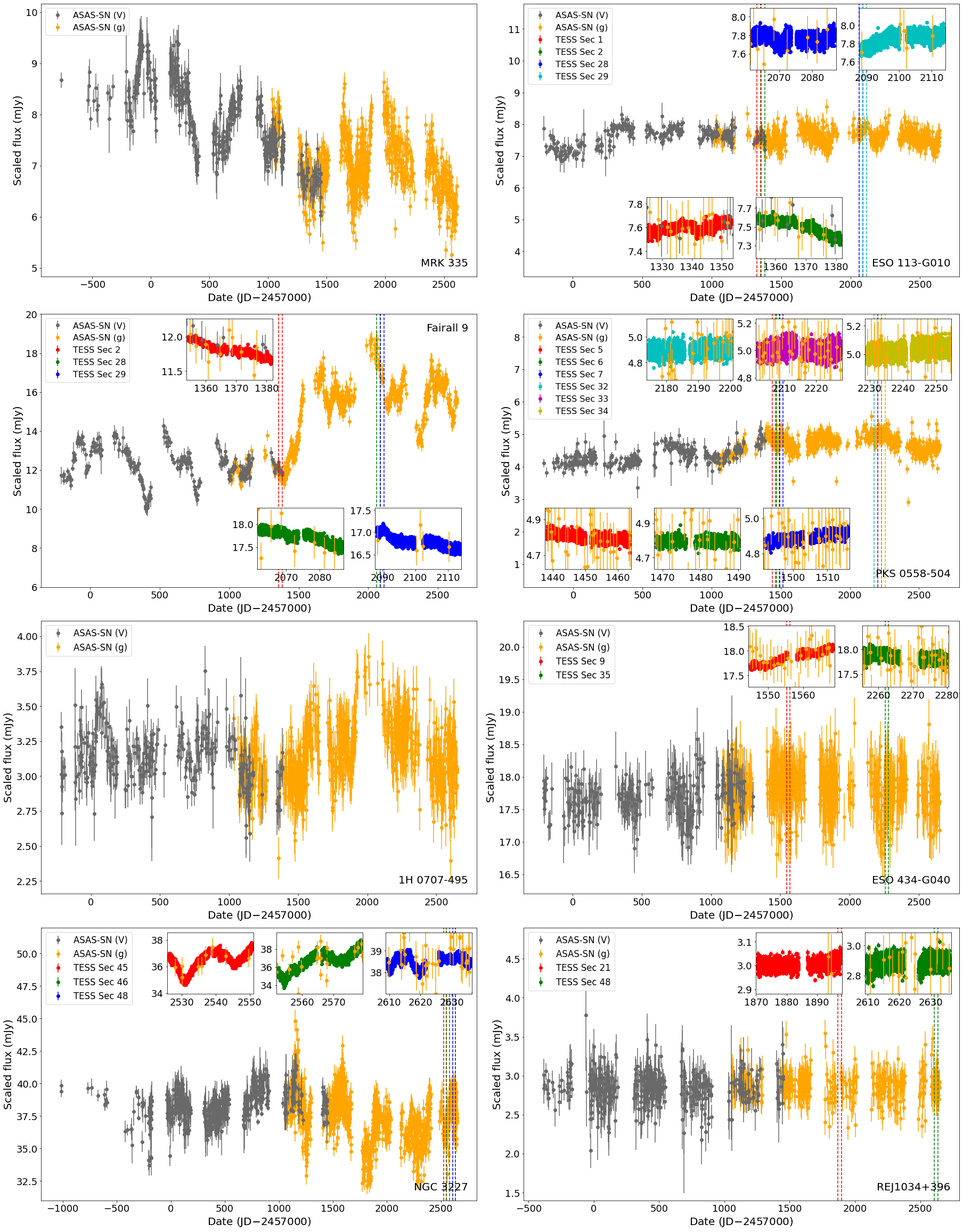}
    \caption{The ASAS-SN light curves of MRK 335, ESO 113-G010, Fairall 9, PKS 0558-504, 1H 0707-495, ESO 434-G040, NGC 3227, and RE J1034+396. with the TESS sectors marked and shown in the insets. The light curves are scaled to overlap.}
    \label{lcs_compilation_1}
\end{figure*}

\begin{figure*}[h!]
    \centering
    \includegraphics[width=0.95\hsize]{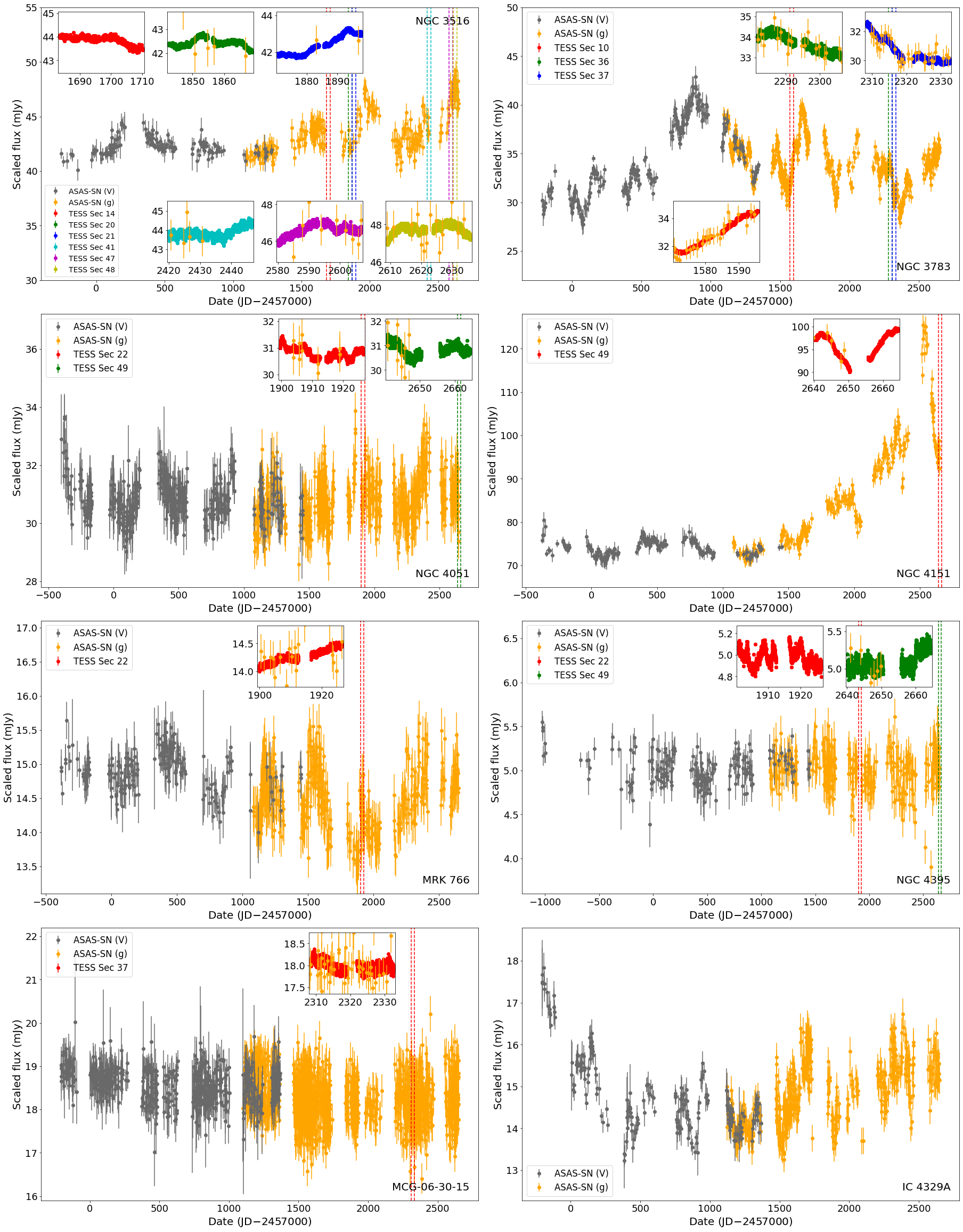}
    \caption{Continuation of Figure \ref{lcs_compilation_1}, for NGC 3516, NGC 3783, NGC 4051, NGC 4151, MRK 766, NGC 4395, MCG-06-30-15, and IC 4329A.}
    \label{lcs_compilation_2}
\end{figure*}

\begin{figure*}[h!]
    \centering
    \includegraphics[width=0.95\hsize]{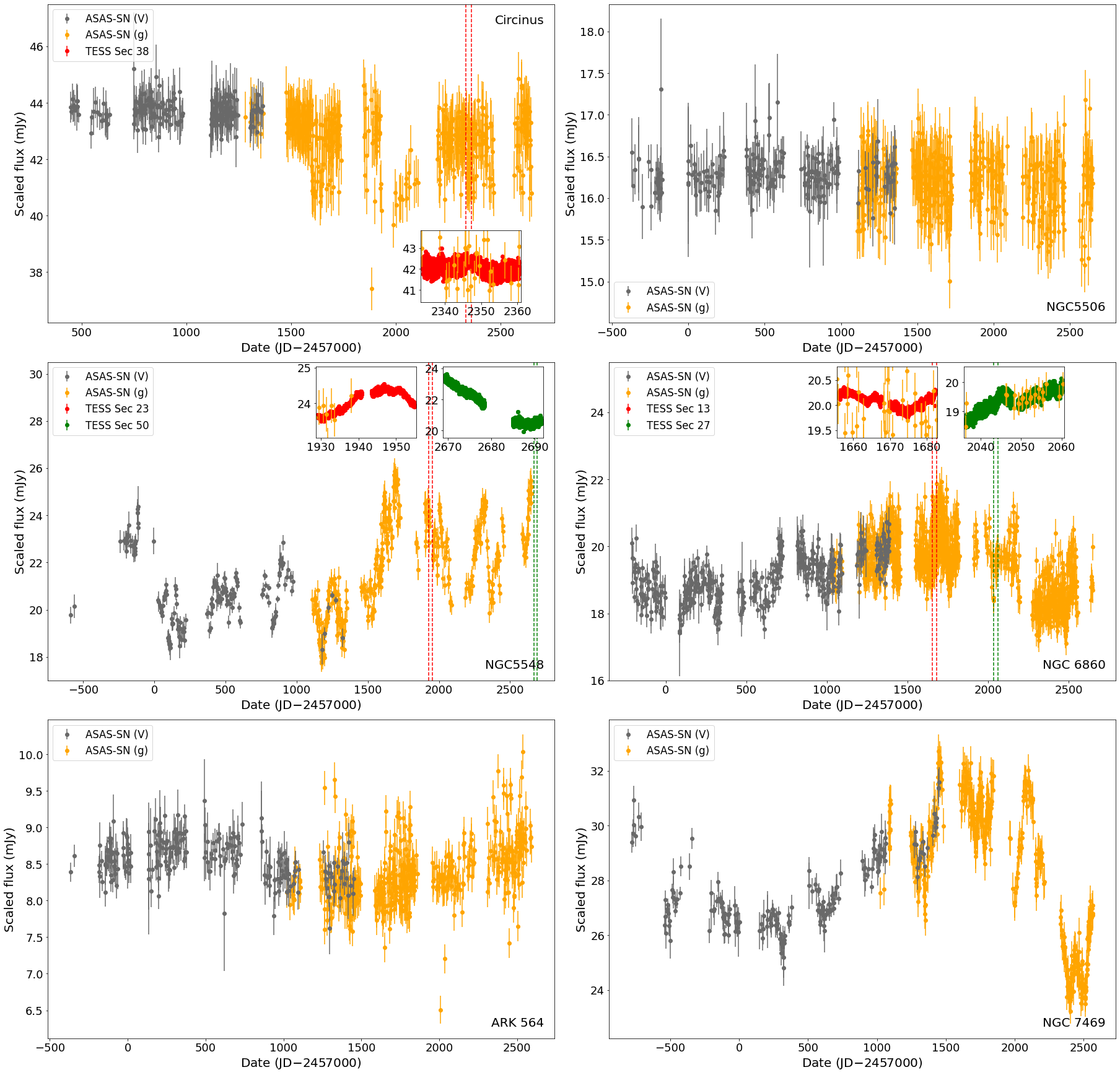}
    \caption{Continuation of Figure \ref{lcs_compilation_1}, for Circinus, NGC 5506, NGC 5548, NGC 6860, ARK 564, and NGC 7469.}
    \label{lcs_compilation_3}
\end{figure*}

\section{Multi-variable correlation fitting results}
\label{secmulti}

\begin{figure*}[h!]
    \centering
    \includegraphics[width=0.9\hsize]{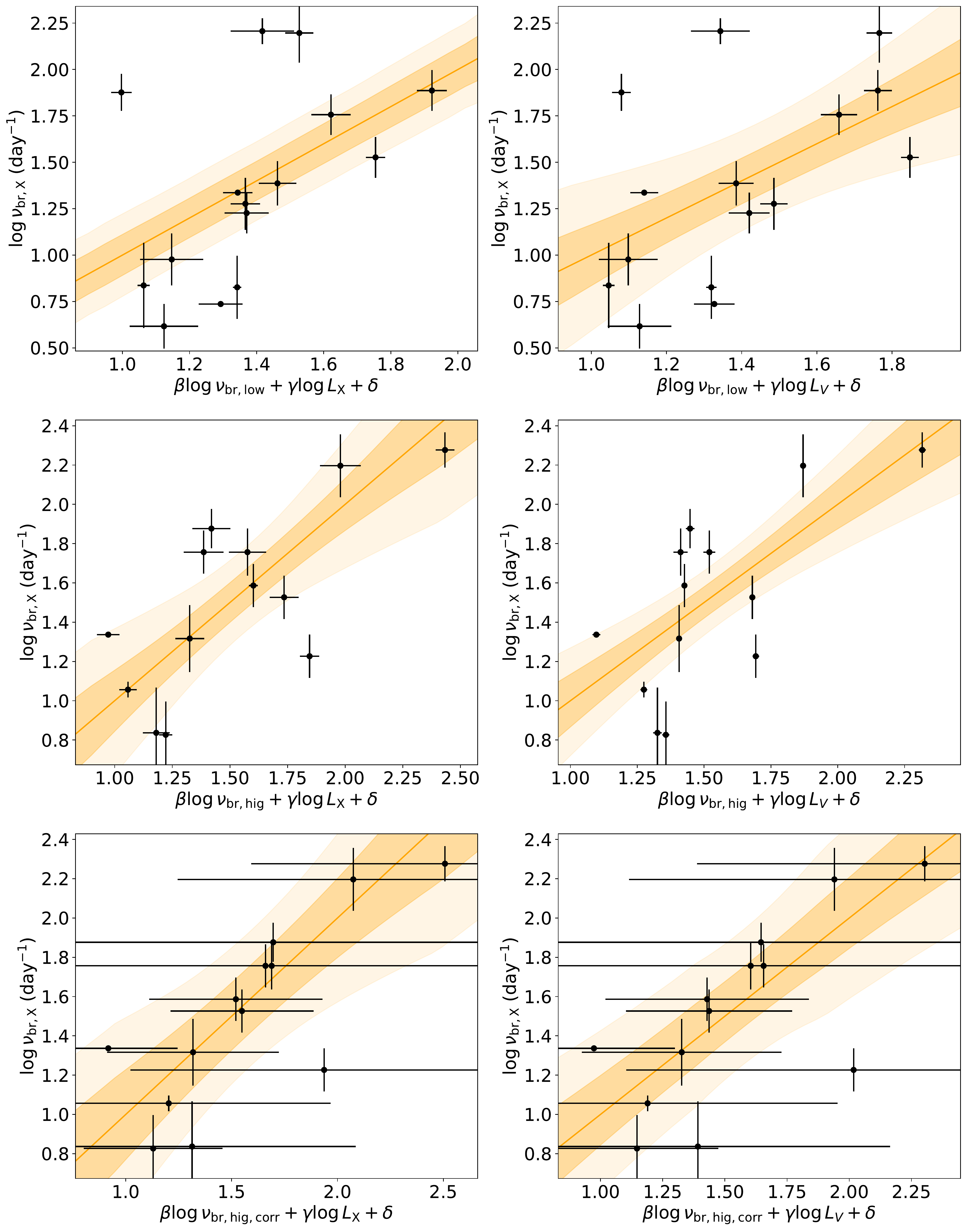}
    \caption{The multi-variable correlation between X-ray break frequency and different optical break frequencies ($\log\nu_{\textrm{br,low}}$ (top), $\log\nu_{\textrm{br,hig}}$ (middle), and $\log\nu_{\textrm{br,hig,corr}}$ (bottom)), with the addition of X-ray luminosity (left) or $V$-band luminosity (right). The dark and lightly shaded regions represent the 68\% and 95\% confidence ranges, respectively.}
    \label{multiplot_1}
\end{figure*}

\begin{figure*}[h!]
    \centering
    \includegraphics[width=0.9\hsize]{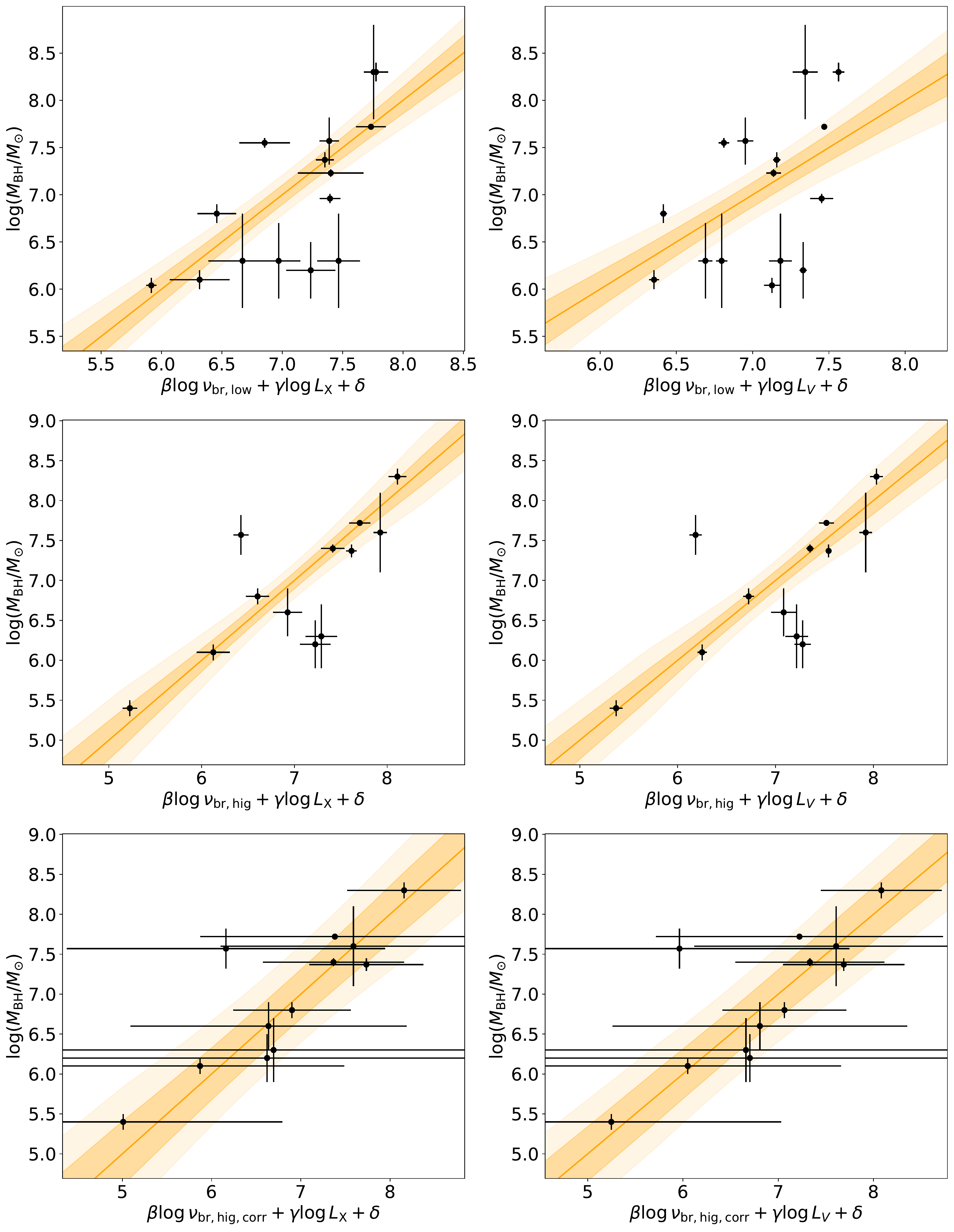}
    \caption{Same as Figure \ref{multiplot_1}, but with $M_\textrm{BH}$ instead of $\nu_{\textrm{br,X}}$.}
    \label{multiplot_2}
\end{figure*}

In this section, we present the results of fitting the multi-variable correlations (Equation \ref{multieq}). Figures \ref{multiplot_1} and \ref{multiplot_2} show the fitting plots with the addition of X-ray and V-band luminosities, which improved the fit the most. Table \ref{multitable} lists all values we obtained from the fits. The units used for $\nu_{\textrm{br}}$, $M_{\textrm{BH}}$, $N_H$, and $L$ are day$^{-1}$, $M_{\odot}$, cm$^{-2}$, and erg s$^{-1}$, respectively.

\begin{table*}[h!]
\caption{Parameters obtained from multi-variable correlation fittings to Equation \ref{multieq} and Pearson correlation strength and null probability. $\beta$ values with no uncertainties indicate the cases where $\beta$ was fixed.}
\label{multitable}
\centering
\begin{tabular}{l c c c c c c c}
\hline \hline
$Y$ & $\log \nu_{\textrm{br,opt}}$ & $X$ & $\beta$ & $\gamma$ & $\delta$ & $r$ & $p$ \\
\hline
$\log \nu_{\textrm{br,X}}$ & $\log \nu_{\textrm{br,low}}$ & $\log N_H$ & $0.83^{+0.05}_{-0.05}$ & $-0.12^{+0.07}_{-0.07}$ & $6.07^{+1.42}_{-1.44}$ & 0.50 & 0.06 \\
 &  & $\log L_\textrm{bol}$ & $0.63^{+0.05}_{-0.05}$ & $0.01^{+0.02}_{-0.02}$ & $2.48^{+0.81}_{-0.82}$ & 0.50 & 0.06 \\
 &  & $\log L_\textrm{X}$ & $0.63^{+0.05}_{-0.05}$ & $0.02^{+0.02}_{-0.02}$ & $2.22^{+0.73}_{-0.73}$ & 0.50 & 0.06 \\
 &  & $\log L_V$ & $0.52^{+0.05}_{-0.05}$ & $-0.19^{+0.06}_{-0.06}$ & $11.13^{+2.59}_{-2.59}$ & 0.54 & 0.04 \\
 &  & $\log (L_V/L_{\textrm{X}})$ & $0.74^{+0.05}_{-0.05}$ & $-0.18^{+0.08}_{-0.08}$ & $3.40^{+0.17}_{-0.17}$ & 0.53 & 0.04 \\
 & $\log \nu_{\textrm{br,hig}}$ & $\log N_H$ & $0.65^{+0.04}_{-0.03}$ & $-0.47^{+0.11}_{-0.11}$ & $12.48^{+2.29}_{-2.42}$ & 0.63 & 0.02 \\
 &  & $\log L_\textrm{bol}$ & $0.57$ & $-0.18^{+0.05}_{-0.05}$ & $10.25^{+2.09}_{-2.1}$ & 0.73 & 0.005 \\
 &  & $\log L_\textrm{X}$ & $0.57$ & $-0.21^{+0.05}_{-0.05}$ & $11.25^{+2.24}_{-2.26}$ & 0.73 & 0.005 \\
 &  & $\log L_V$ & $0.21^{+0.04}_{-0.04}$ & $-0.25^{+0.05}_{-0.05}$ & $12.75^{+2.28}_{-2.29}$ & 0.69 & 0.009 \\
 &  & $\log (L_V/L_{\textrm{X}})$ & $0.66^{+0.04}_{-0.04}$ & $0.60^{+0.09}_{-0.09}$ & $1.70^{+0.13}_{-0.13}$ & 0.69 & 0.01 \\
 & $\log \nu_{\textrm{br,hig,corr}}$ & $\log N_H$ & $0.18^{+0.06}_{-0.06}$ & $-0.72^{+0.21}_{-0.21}$ & $17.08^{+4.49}_{-4.56}$ & 0.59 & 0.03 \\
 &  & $\log L_\textrm{bol}$ & $0.20$ & $-0.18^{+0.06}_{-0.06}$ & $9.33^{+2.45}_{-2.49}$ & 0.78 & 0.002 \\
 &  & $\log L_\textrm{X}$ & $0.20$ & $-0.21^{+0.06}_{-0.06}$ & $10.43^{+2.56}_{-2.59}$ & 0.78 & 0.002 \\
 &  & $\log L_V$ & $0.20$ & $-0.12^{+0.04}_{-0.04}$ & $6.66^{+1.68}_{-1.57}$ & 0.70 & 0.008 \\
 &  & $\log (L_V/L_{\textrm{X}})$ & $0.11^{+0.03}_{-0.03}$ & $0.65^{+0.09}_{-0.1}$ & $0.79^{+0.15}_{-0.16}$ & 0.71 & 0.006 \\ \hline
$\log M_{\textrm{BH}}$ & $\log \nu_{\textrm{br,low}}$ & $\log N_H$ & $-0.86^{+0.09}_{-0.09}$ & $0.1^{+0.04}_{-0.04}$ & $2.74^{+0.86}_{-0.85}$ & 0.39 & 0.15 \\
 &  & $\log L_\textrm{bol}$ & $-0.30^{+0.09}_{-0.09}$ & $0.53^{+0.06}_{-0.06}$ & $-16.62^{+2.42}_{-2.43}$ & 0.71 & 0.003 \\
 &  & $\log L_\textrm{X}$ & $-0.30^{+0.09}_{-0.09}$ & $0.62^{+0.06}_{-0.06}$ & $-19.81^{+2.7}_{-2.7}$ & 0.71 & 0.003 \\
 &  & $\log L_V$ & $-0.50^{+0.09}_{-0.09}$ & $0.45^{+0.07}_{-0.07}$ & $-13.94^{+3.01}_{-2.92}$ & 0.49 & 0.07 \\
 &  & $\log (L_V/L_{\textrm{X}})$ & $-0.52^{+0.09}_{-0.09}$ & $-0.58^{+0.08}_{-0.08}$ & $6.52^{+0.26}_{-0.26}$ & 0.60 & 0.02 \\
 & $\log \nu_{\textrm{br,hig}}$ & $\log N_H$ & $-1.60^{+0.08}_{-0.08}$ & $1.18^{+0.18}_{-0.18}$ & $-19.92^{+3.85}_{-3.82}$ & 0.61 & 0.03 \\
 &  & $\log L_\textrm{bol}$ & $-1.09$ & $0.38^{+0.05}_{-0.05}$ & $-10.78^{+2.02}_{-2.02}$ & 0.77 & 0.003 \\
 &  & $\log L_\textrm{X}$ & $-1.09$ & $0.43^{+0.05}_{-0.05}$ & $-12.72^{+2.26}_{-2.28}$ & 0.77 & 0.004 \\
 &  & $\log L_V$ & $-1.09$ & $0.33^{+0.05}_{-0.05}$ & $-8.61^{+2.0}_{-1.99}$ & 0.70 & 0.01 \\
 &  & $\log (L_V/L_{\textrm{X}})$ & $-1.29^{+0.08}_{-0.08}$ & $-1.08^{+0.1}_{-0.1}$ & $6.45^{+0.18}_{-0.18}$ & 0.64 & 0.02 \\
 & $\log \nu_{\textrm{br,hig,corr}}$ & $\log N_H$ & $-0.45^{+0.13}_{-0.13}$ & $1.63^{+0.18}_{-0.19}$ & $-27.93^{+3.99}_{-3.93}$ & 0.66 & 0.02 \\
 &  & $\log L_\textrm{bol}$ & $-0.39$ & $0.36^{+0.06}_{-0.06}$ & $-9.01^{+2.4}_{-2.39}$ & 0.84 & 0.0006 \\
 &  & $\log L_\textrm{X}$ & $-0.39$ & $0.42^{+0.06}_{-0.06}$ & $-11.00^{+2.7}_{-2.69}$ & 0.84 & 0.0007 \\
 &  & $\log L_V$ & $-0.39$ & $0.29^{+0.04}_{-0.04}$ & $-5.88^{+1.77}_{-1.76}$ & 0.77 & 0.003 \\
 &  & $\log (L_V/L_{\textrm{X}})$ & $-0.26^{+0.06}_{-0.06}$ & $-1.07^{+0.1}_{-0.1}$ & $8.10^{+0.25}_{-0.25}$ & 0.74 & 0.006 \\
\hline
\end{tabular}
\end{table*}

\end{appendix}

\end{document}